\documentclass[twoside,11pt]{article}
\usepackage{jair, theapa, rawfonts}

\jairheading{X}{XXXX}{X-XX}{X/XX}{X/XX}

\usepackage{soul}
\usepackage[utf8]{inputenc}
\usepackage[small]{caption}
\usepackage{graphicx}
\usepackage{amsmath}
\usepackage{amsthm}
\usepackage{booktabs}
\usepackage{algorithm}
\usepackage{algorithmic}

\usepackage{lipsum}  
\usepackage{amssymb}
\newcommand{\sol}{\text{SOL}}
\newcommand{\opt}{\text{OPT}}
\newcommand{\cms}{\textsc{cms}}
\newcommand{\msat}{\textsc{min} \textsc{2-sat}}

\usepackage{graphicx}
\newcommand{\bigoh}{\mathcal{O}}
\newcommand{\np}{NP}

\usepackage{mathtools}

\usepackage{diagbox}
\newcommand{\PreserveBackslash}[1]{\let\temp=\\#1\let\\=\temp}
\newcolumntype{C}[1]{>{\PreserveBackslash\centering}p{#1}}

\usepackage{caption}
\usepackage{subcaption}
\newcommand\unaryminus{\smash{\scalebox{0.35}[1.0]{\( - \)}}}
\usepackage{adjustbox}
\usepackage{enumitem}
\newenvironment{Figure}
  {\par\medskip\noindent\minipage{\linewidth}}
  {\endminipage\par\medskip}
\usepackage{amsthm}
\usepackage{mathtools}
\usepackage[switch]{lineno} 
\newtheorem{theorem}{Theorem}
\newtheorem{corollary}{Corollary}
\newtheorem{lemma}[theorem]{Lemma}

\newtheorem{proposition}{Proposition}
\newtheorem{example}{Example}
\newtheorem{claim}{Claim}
\theoremstyle{definition}
\newtheorem{definition}{Definition}
\newtheorem{remark}{Remark}

\newcommand{\csp}{\textsc{csp}}
\newcommand{\csps}{\textsc{csp}\!\,s}
\newcommand{\cavone}{\textsc{cav-1}}
\newcommand{\cdvone}{\textsc{cdv-1}}
\newcommand{\caaone}{\textsc{caa-1}}
\newcommand{\cdaone}{\textsc{cda-1}}
\newcommand{\cavall}{\textsc{cav-all}}
\newcommand{\cdvall}{\textsc{cdv-all}}
\newcommand{\caaall}{\textsc{caa-all}}
\newcommand{\cdaall}{\textsc{cda-all}}
\newcommand*\xor{\mathbin{\oplus}}

\usepackage[table,xcdraw]{xcolor}
\usepackage[implicit=false]{hyperref}
\usepackage{physics}
\usepackage{amsmath}
\usepackage{tikz}
\usepackage{mathdots}
\usepackage{yhmath}
\usepackage{cancel}
\usepackage{color}
\usepackage{siunitx}
\usepackage{array}
\usepackage{multirow}
\usepackage{amssymb}
\usepackage{tabularx}
\usepackage{booktabs}
\usetikzlibrary{fadings}
\usetikzlibrary{patterns}
\usetikzlibrary{shadows.blur}
\usetikzlibrary{shapes}
\usepackage{adjustbox}
\usepackage{multicol}
\usepackage{cleveref}[2012/02/15]
\crefformat{footnote}{#2\footnotemark[#1]#3}

\let\ACMmaketitle=\maketitle
\renewcommand{\maketitle}{\begingroup\let\footnote=\thanks \ACMmaketitle\endgroup}

\newenvironment{proofof}[1]
      {\smallskip\noindent{\bf #1.}}
      {\hfill$\Box$\medskip}

\usepackage{bold-extra}

\begin{document}

\title{On the Complexity of Winner Determination and \\Strategic Control in Conditional Approval Voting\footnote{This work has evolved as an extended merge of two preliminary conference publications: \cite{MP20,MP21}.}
\vspace{-0.4cm}
}

\author{\name Evangelos Markakis \email markakis@aueb.gr\\ 
\addr Athens University of Economics and Business, Greece\\ Archimedes, Athena Research Center, Greece\\ Input Output Global (IOG), Greece.
\AND
       \name Georgios Papasotiropoulos \email gpapasotiropoulos@aueb.gr \\
       \addr University of Warsaw, Poland \& Athens University of Economics and Business, Greece  
      }

\maketitle

\vspace{-0.4cm}
{\small{
\begin{abstract}
We focus on a generalization of the classic Minisum
approval voting rule, introduced by Barrot
and Lang (2016), and referred to as Conditional
Minisum (\cms), for multi-issue elections with preferential dependencies.
Under this rule, voters are allowed to declare dependencies between different issues, but the price we have to pay for this higher level of expressiveness is that we end up with a computationally hard rule. Motivated by this, we first focus on finding special cases that admit efficient algorithms for \cms. Our main result in this direction is that we identify the condition of bounded treewidth (of an appropriate graph, emerging
from the provided ballots) as the necessary and sufficient
condition for exact polynomial algorithms, under
common complexity assumptions. 
We then  move to the design of approximation algorithms. For the (still hard) case of binary issues, we identify natural restrictions on the voters' ballots, under which we provide the first multiplicative approximation algorithms for the problem.
The restrictions involve upper bounds on the number of dependencies an issue can have on the others and on the number of alternatives per issue that a voter can approve. 
Finally, we also investigate the complexity of 
problems related to the strategic control of conditional approval elections by adding or deleting either voters or alternatives and we show that in most variants of these problems, \cms\ is computationally resistant against control. Overall, we conclude that \cms\ can be viewed as a solution that achieves a satisfactory tradeoff between expressiveness and computational efficiency, when we have a limited number of dependencies among issues, while at the same time exhibiting sufficient resistance to control.
\end{abstract}}}



\vspace{-0.3cm}
\section{Introduction}

Over the years, the field of social choice theory has focused more and more on decision making over combinatorial domains \cite{LX16}, which involves settings like \textit{multi-winner elections}, e.g. for the formation of a committee, and elections for a set of issues that need to be decided upon simultaneously, often referred to as {\it multiple referenda}.
In this work, we focus on approval voting as a means for collective decision making on multiple issues with multiple alternatives each. Approval voting offers a simple and easy to use format for running such elections, by having voters express an approval or disapproval separately for the alternatives of each issue. 
There is already a range of voting rules that are based on approval ballots, including the classic Minisum solution, which for each issue selects the alternative with the highest support from the electorate, along with more recently introduced methods as outlined in the ``Related Work" section, below. 

However, the rules most commonly studied for approval voting are applicable only when the issues under consideration are independent. 
As soon as the voters exhibit preferential dependencies between the issues, we have more challenges to handle. More precisely, voters' preferences on a specific issue may be conditioned upon the outcome of other issue(s) and
this is not uncommon in practical scenarios:
A resident of a municipality may wish to support public project A, only if public project B is also implemented (which she evaluates as more important); a faculty member may want to vote in favor of hiring a new colleague only if the other new hires have a different research expertise; a group of friends may want to go to a certain movie theater only if they decide to have dinner at a nearby location; festival organizers could choose to approve the inclusion of several musical acts in their lineup but decide to limit the number of acts to a small fixed number, e.g., due to budget constraints; a grant committee may approve funding for Project X, but only if Project Y didn't receive sufficient support from the committee members to be implemented. 
We can also consider another example with conditional preferences, taken from recommendation systems for online advertising: suppose an ad management service needs to make a personalized selection of ads, to be shown on Alice's favorite news website. For each slot (or area) in the advertising region of the site, there is a set of possible ads to choose from and the overall goal is to maximize the likelihood that Alice will click on one of these ads. Her likelihood to click depends on whether she encounters ads that strongly align with her interests. If we think of the slots as corresponding to issues, a recommendation could be made by looking at the data from users ``similar" to Alice (voters), and their clicking behavior (approvals). Notably, these voters have conditional preferences, as their probability of clicking on an ad is influenced when a related ad appears in a nearby slot. For instance, the probability may increase for products frequently bought together or decrease when a product is defamed by the ad of another

It is rather obvious that voting separately for each issue cannot provide a good solution in any of the 
above settings. 
Consequently, as detailed in the ``Related Work" section, several approaches have been suggested to take into account preferential dependencies. Nevertheless, the majority of these works are suitable for rules where voters are required to express a ranking over the set of issues or have a numerical representation of their preferences instead of approval-based preferences. 
    The first work that introduced a framework for expressing dependencies exclusively in the context of approval voting was by Barrot and Lang \citeyear{BL16}. They defined the notion of a conditional approval ballot (where the voters can specify a dependency graph for the issues of the election in conjunction with their ballots) and introduced new voting rules, that generalized some of the known rules from the literature of the standard approval setting. Among the properties that were studied, it was also exhibited that, in general, a higher level of expressiveness implies higher computational complexity. More precisely, the Minisum solution (also frequently referred to as the Approval Voting rule) is known to be efficiently computable in the standard (unconditional) approval setting, but its generalization, referred to as {\it Conditional Minisum} (or \cms\ in short), was shown to be \np-hard. In the unconditional approval setting, the Minisum solution stands as the most straightforward method for selecting winning alternatives, and it has established itself as one of the primary election systems extensively examined in economic theory, political science and computational social choice, being also widely used in practice \cite{ulle}. 
    
    Given how central the Minisum solution is, and how practical conditional ballots can be in real-life scenarios, and in light of the computational challenges presented by \citeauthor{BL16} \citeyear{BL16}, it becomes natural to investigate whether \cms\ admits exact algorithms for certain families of instances or approximation algorithms with provable guarantees. 
Progress on this front would allow us to draw conclusions on the applicability of approval-based elections in which voters are endowed with a significant degree of expressiveness.

In addition to our paper's focus on winner determination under \cms\ rule, we also take a step towards examining potential threats, that could arise due to malicious behaviour within elections over interdependent issues. These explorations can, in principle, contribute to enhancing the fairness and transparency of elections, ensuring the overall integrity of the procedure, and paving the way for the development of algorithms aimed at detecting and preventing malicious attempts. In the realm of strategic considerations, the primary focal points within the computational social choice literature revolve around questions of strategyproofness and election control. Strategyproofness is the axiom that is met when no voter can increase her satisfaction with respect to the rule's outcome by misreporting her true preferences; in contrast to the unconditional case, \cms\ is known to be manipulable \cite{BL16}. In our work, we focus on elections' control and we study a spectrum of scenarios where the election conductor seeks to control the election outcome, so as to align with their own preferences, through various strategic actions. While, in many instances, a controller may be able to influence the input of the election so as to successfully enforce her will, we examine, from a computational complexity perspective, the question of whether the conductor can always and in polynomial time exert control over the outcome. To be more precise, we narrow our focus to worst-case scenarios, examining the existence of instances where the conductor encounters inherent computational challenges in achieving their desired outcomes. Similar questions form a very prominent research agenda within computational social choice as it pertains to understanding the susceptibility of election systems. 

Quoting \citeauthor{HHR09} \citeyear{HHR09}, the ``dream case" of a voting rule is one that features an efficient algorithm for determining the winning candidate(s), while making any form of strategic control by malicious external agents computationally infeasible. Motivated by this, we aimed to explore, from an algorithmic perspective, both the efficient computation (exact and approximate) of the winner determination problem and the computational complexity related to aspects of election control. Overall, our work indicates that \cms\ stands as a strong candidate for real-life applications as it strikes a favorable balance between voters' expressiveness and computational efficiency, all while exhibiting sufficient (computational) resilience against most malicious efforts.

\paragraph{Contribution.} We undertake a study of the Conditional Minisum voting rule, a.k.a. \cms, which attempts to minimize the total dissatisfaction score across all voters in conditional approval elections, from the viewpoint of algorithms and complexity. Our goals are:
\begin{enumerate}
    \item[(a.)] to enhance the understanding on the complexity implications due to conditional voting for the winner determination problem under a rule that is known to be efficiently computable in the absence of dependencies between issues, with respect to both exact and approximate solutions,
    \item[(b.)] to identify whether the rule is immune to, or, at least, computationally resistant against malicious control by strategic actions.
\end{enumerate} 

In Section \ref{sec:opt}, we focus on conditions that lead to exact polynomial time algorithms for computing the optimal solution under \cms\ voting rule. For this, we consider the intuitively simple (but still \np-hard) case, where each issue can depend on at most one other issue for every voter, and our main insight is that one can draw conclusions by looking at (undirected variant of) the global dependency graph of an instance, which is formed by taking the union of dependencies by all voters. We later generalize this result for dependencies on any constant number of issues. Restrictions on the structure of the global dependency graph allow us to identify the condition of bounded treewidth as the only restriction that leads to optimal efficient algorithms. More precisely, our results provide characterizations for the families of \cms\ instances that can be placed in P and FPT, implying that the condition of bounded treewidth serves as the lynchpin between expressiveness of voters' ballots and efficiency of solving the winner determination problem. These results also establish a connection with well studied classes of Constraint Satisfaction Problems, which can be of independent interest. 

In Section \ref{sec:approx}, we provide the first multiplicative approximation algorithms for conditional approval elections of issues with binary domain, under the condition that for every voter, each issue can depend on at most one other issue. The considered family of instances, includes the set of instances that were proven to be \np-hard by \citeauthor{BL16} \citeyear{BL16}. In the corresponding graph-theoretic representation of the problem, which will be introduced in Section \ref{sec:prelims}, the condition corresponds to voters with dependency graph of maximum in-degree no more than 1. The main positive result of the section is an algorithm that achieves an approximation factor of $1.1037$. Interestingly, our algorithm is based on a reduction to \textsc{min sat}, an optimization version of \textsc{sat} that has rarely been applied in computational social choice (in contrast to \textsc{max sat}). The result is contingent upon an additional, but well-motivated from the perspective of social choice, assumption regarding the number of approved alternatives per issue in a voter's ballot. Imposing such a further requirement might at first seem demanding, however, we have also established a strong negative result: in the absence of further assumptions, no algorithm can attain any bounded multiplicative approximation guarantee, even for instances with binary domains and even if for every voter, each issue depends on at most one other issue. Concluding the section, we put forth some additional (and similar in flavor) assumptions that, when satisfied, also enable the existence of provable approximation guarantees, albeit non-constant. Interestingly, these results also relate the considered voting rule with classical algorithmic problems.

Moving on, in Section \ref{sec:controlling}, we initiate the algorithmic study of some standard notions of election control for \cms. The problems we examine concern the attempt by an external agent to enforce the election of certain alternative(s) in either one or every issue under consideration, by adding or deleting either voters or alternatives. We consider a total of 8 variants of this question, depending on the number of issues to be controlled and on whether we have addition or deletion of voters/alternatives. We provide a set of 18 computational complexity results that give a complete picture with respect to the crucial parameters of the input in every one of the considered problems. Our findings reveal that \cms\ is sufficiently computationally resistant, against such moves.

\paragraph{Related Work.} 
Approval voting for multi-issue elections has gained great attention in the recent years, driven by its simplicity and practical potential. 
Apart from the classic Minisum solution \cite{weber78,brams78,brams07,laslier10}, other rules have also been considered, such as the Minimax solution \cite{BKS07}, Satisfaction Approval Voting \cite{BK10}, families of rules based on Weighted Averaging Aggregation \cite{Amanatidis+15}, Proportional Approval Voting and Chamberlin-Courant. The last two rules, as well as Minisum, can be captured by the general family of Thiele voting rules; for these (and other approval based rules) we refer to the recently published book by \citeauthor{lackner2023multi} \citeyear{lackner2023multi} and to the surveys by \citeauthor{BF05} \citeyear{BF05} and \citeauthor{Kilgour10} \citeyear{Kilgour10}. None of these rules however allow voters to express dependencies. 
The first work that exclusively took this direction for approval-based elections is by Barrot and Lang \citeyear{BL16}. Namely, three voting rules were proposed for incorporating such dependencies (including the Conditional Minisum rule that we consider here) and some of their properties were studied, mainly on the satisfiability of certain axioms. Conditional approval ballots have a clear resemblance with the well-studied model of CP-nets \cite{boutilier2004cp}, which is a graphical representation of voters' preferences depicting conditional dependence and independence of preference statements under a ceteris paribus (all else being equal) interpretation, but, as it has been highlighted in the work of \citeauthor{BL16} \citeyear{BL16} the two frameworks define different semantics and are incomparable. 

Even if one moves away from approval-based elections, the presence of preferential dependencies remains a major challenge when voting over combinatorial domains. Several methodologies have been considered 
achieving various levels of trade-offs between expressiveness and efficient computation. Some representative examples include, among others, sequential voting \cite{LX09,Airiau+11,dallaPozza+11,CX12}, or completion principles for partial preferences \cite{LL09,CL12}. Analogous attempts to increase the expressiveness of agents' ballots have been also examined in other subfields of computational social choice; indicatively we refer to Participatory Budgeting \cite{rey23,jain20}, Judgement Aggregation \cite{grandi10} and Liquid Democracy \cite{colley21}.

Finally, we consider versions of control that fall within the standard approaches that have been used for studying the complexity of affecting election outcomes. For an extensive study on this topic, we refer to the work of \citeauthor{FRM16} \citeyear{FRM16}. Indicatively, the study of adding or deleting voters or alternatives began with the paper of \citeauthor{BTT92} \citeyear{BTT92} and some subsequent works are by \citeauthor{gupta2022resolute} \citeyear{gupta2022resolute}, \citeauthor{yang2023complexity} \citeyear{yang2023complexity},
\citeauthor{HHR07} \citeyear{HHR07},
\citeauthor{FHL11} \citeyear{FHL11},
\citeauthor{Liu+09} \citeyear{Liu+09},
\citeauthor{meir2008complexity} \citeyear{meir2008complexity},
\citeauthor{bulteau2015combinatorial} \citeyear{bulteau2015combinatorial}, and \citeauthor{kellerhals2017computational} \citeyear{kellerhals2017computational}.

\section{Formal Background}
\label{sec:prelims}

Let $I = \{I_1,\dots,I_m\}$ be a set of $m$ issues, where each issue $I_j$ is associated with a finite domain $D_j$ of alternatives. An {\it{outcome}} is an assignment of a value for every issue, and let $D = D_1 \times D_2 \times \dots \times D_m$ be the set of all possible outcomes. Let also $V = \{1,\dots, n\}$ be a group of $n$ voters who have to decide on a common outcome from $D$.

\paragraph{Voting Format.} To express dependencies among issues, we mostly follow the format described in the paper by \citeauthor{BL16} \citeyear{BL16}. Each voter $i\in[n]$ is associated with a directed graph $G_i=(I,E_i)$, called {\it{dependency graph}}, whose vertex set coincides with the set of issues. A directed edge $(I_k, I_j)$ means that issue $I_j$ is affected by $I_k$. We also let $N^{\unaryminus}_{i}(I_j)$ be the (possibly empty) set of direct predecessors of issue $I_j$ in $G_i$.
We first explain briefly how the voters are expected to submit their preferences, before giving the formal definition. For an issue $I_j$ that has no predecessors in $G_i$ (in other words, its in-degree is 0), voter $i$ is allowed to cast an unconditional approval ballot, stating the alternatives of $D_j$ that are approved by her. In the case that issue $I_j$ has a positive in-degree in $G_i$, then let $\{I_{j_1}, I_{j_2}, \dots, I_{j_k} \}\subseteq I$ be all its direct predecessors (also called in-neighbors).
Voter $i$ then needs to specify all the combinations that she approves in the form $\{t:r\}$ where $r\in D_j$, and $t \in D_{j_1} \times D_{j_2} \times \dots \times D_{j_k}$. Every such combination $\{t:r\}$ signifies the satisfaction of voter $i$ with respect to issue $I_j$ in a given outcome, when that outcome contains all alternatives in $t$ as well as the alternative $r$ for the issue $I_j$. Both cases of zero and positive in-degree for an issue can be unified in the following definition of conditional approval ballots.

\begin{definition}
	A conditional approval ballot of a voter $i$ over issues $I=\{I_1,\dots,I_m\}$ with domains $D_1,\dots, D_m$ respectively, is a pair $B_i=\langle G_i,\{A_j, j\in [m]\} \rangle$, where $G_i$ is the dependency graph of voter $i$, and for each issue $I_j$, $A_j$ is a set of conditional approval statements in the form $\{t: r\}$ with $t\in \prod_{k\in N^{\unaryminus}_{i}(I_j)} D_k$, and $r \in D_j$. 	
\end{definition}

 To simplify the presentation, when a voter has expressed a common dependency for $k>1$ alternatives of an issue $I_j$, we can group them together and write $\{t:\{d_j^1, d_j^2, \dots, d_j^k\}\}$, instead of $\{t: d_j^1\}$, $\{t: d_j^2\}$, $\dots$, $\{t: d_j^k\}$.
Additionally, for every issue $I_j$ with in-degree $0$ by some voter $i$, a vote in favor of $d_j$ will be written simply as $\{d_j\}$, instead of $\{\emptyset: d_j\}$.

An important quantity for parameterizing families of instances is the maximum in-degree\footnote{When $\Delta_i$ is large for some voter $i$, the input might become exponentially large.
Alternatively, one could try a succinct way of representing ballots using propositional formulae. We will not examine further this issue, since for the cases that we consider, the in-degree is constant.} of each graph $G_i$, namely $\Delta_i=\max_{j \in [m]}\{|N^{\unaryminus}_{i}(I_j)|\}$. Let also $\Delta = \max_{i\in [n]} \Delta_i$. Given a voter $i$ with conditional ballot $B_i$, we will denote by $B_i^j$ the restriction of her ballot to issue $I_j$. Moreover, a \textit{conditional approval voting profile} is given by a tuple $P = (I,D,V,B)$, where $B=( B_1, B_2, \dots, B_n)$.

\begin{definition}
The global dependency graph of a set of voters is the undirected
simple graph that emerges from ignoring the orientation of edges in the graph $(I,\bigcup_{i\in[n]}E_i)$, where $E_i$ is the edge set of the dependency graph of voter $i$. 
\end{definition}

\begin{example}
\normalfont
\label{example:ex1}
	
As an illustration, we consider 3 co-authors of some joint research who have to decide on $3$ issues: whether they will {\textit{work}} (w) more before the submission deadline on obtaining new theorems, whether they have enough material to split their work into two, or even \textit{multiple} (m), papers or submit all their results in a single submission, and whether they should invite a new {\textit{co-author}} (c) to work with them because of his insights that can help on improving their results. The first author insists on more work before the submission, additionally he approves the choice of two submissions if and only if they work more on new theorems. Furthermore, he does not want to have a new co-author if and only if they split their work. The second author does not have time for more work before the deadline, he has no strong opinion on multiple submissions, approving both alternatives, and he agrees with inviting a new co-author only if they decide both to work more for new results and to submit a single paper. Finally, the last author is interested in working more and in splitting their work and she does not have a strong opinion on whether she prefers to invite a new co-author or not, unless they all decide not to work more neither to make more than a single submission, in which case she disagrees with such an invitation. 

More formally, let $I=\{I_1,I_2,I_3\}$ be the aforementioned issues where $D_1=\{w,\overline{w}\},$ $ D_2=\{m,\overline{m}\},D_3=\{c,\overline{c}\}$. The dependency graphs and the voters' preferences follow.

	\begin{multicols}{2}

\begin{Figure}
\hspace{1.5cm}
\includegraphics[width=1.35\linewidth]{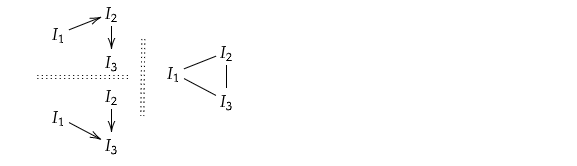}
\captionof{figure}{The dependency graph of voter 1 (up left), voters 2 and 3 (down left) and the global dependency graph (right).}
\end{Figure}
	\columnbreak

	\begin{table}[H]
  \hskip1cm\begin{tabular}{c|c|c}
			voter 1  & voter 2 & voter 3  \\ \hline
		
			$w$         & 
			  $\{\overline{w},m,\overline{m}\}$ & 
			  $\{w,m\}$
			     \\
			
			$\overline{w}:\overline{m}$ & $wm:\overline{c}$        & $wm:\{c,\overline{c}\}$  \\
			
		$w:m$        & $\overline{w}m:\overline{c}$        &  $\overline{w}m:\{c,\overline{c}\}$      \\
	
			$m:\overline{c}$          & $w\overline{m}:c$        & $w\overline{m}:\{c,\overline{c}\}$   \\
			
			$\overline{m}:c$        & $\overline{w}\overline{m}:\overline{c}$        & \hspace{-0.667cm}$\overline{w}\overline{m}:\overline{c}$ 
		\end{tabular}
  \captionof{figure}{The conditional ballots of the voters.}
	\end{table}
    \end{multicols}

\end{example}

\paragraph{Voting Rule.} In this work, we study a generalization of the classic Minisum solution in the context of conditional approval voting. To do so, we first define a measure for the dissatisfaction of a voter given an assignment of values to all the issues, using the following generalization of Hamming distance.

\begin{definition}
Given an outcome $s=(s_1,s_2,\dots, s_m) \in D$, we say that voter $i$ is dissatisfied (or disagrees) with issue $I_j$, if for the projection of $s$ on $N^{\unaryminus}_i(I_j)$, say $t$, it holds that $\{t : s_j\}\notin B_i^j$. We denote as $\delta_i(s)$ the total number of issues that dissatisfy voter $i$.
\end{definition}

Coming back to Example \ref{example:ex1}, the values of $\delta_i(s)$ for every outcome $s$ and voter $i$ follow.
		
\begin{table}[H]
\centering
	\begin{tabular}{c|cccccccc}
		\multicolumn{1}{l|}{$\delta_{i}(\cdot)$} & \multicolumn{1}{l}{\textit{	\scalebox{.95}{{$wmc$}}}} & \multicolumn{1}{l}{	\scalebox{.95}{{$wm\overline{c}$}}} & 	\scalebox{.95}{{$w\overline{m}c$}} & 	\scalebox{.95}{{$w\overline{m}\overline{c}$}} & 	\scalebox{.95}{{$\overline{w}mc$}} & 	\scalebox{.95}{{$\overline{w}m\overline{c}$}} & 	\scalebox{.95}{{$\overline{w}\overline{m}c$}} & 	\scalebox{.95}{{$\overline{w}\overline{m}\overline{c}$}}  \\ \hline
		\scalebox{.95}{voter $1$}                        & 1                                                 &      0                                             &     1                 &       2               &       3               &         2             &        1              &             2         \\
		\scalebox{.95}{voter $2$}                       & 2                                                &               1                                    &          1            & 2                      &          1            &     0                 &              1        &      0                \\
		\scalebox{.95}{voter $3$}                        & 0                                             &                      0                             &              1        & 1                      &                1      &           1           &                3      &               2             
	\end{tabular}
\end{table}

The rule that our work deals with is {\textit{Conditional Minisum}} (\textsc{cms}) and outputs the outcome that minimizes the total number of disagreements over all voters. To simplify notation, we will use \textsc{cms} to refer both to the voting rule and to the related algorithmic problem; the exact meaning will always be clear from the context.
Formally, the algorithmic problem that we study is as follows. 

\begin{table}[H]
	\centering
	\begin{tabular}{lp{10.5cm}}  
		\toprule
	 \multicolumn{2}{c}{\textsc{conditional minisum (cms)} } \\
		\midrule
\textbf{Given:} & A voting profile $P$ with $m$ issues and $n$ voters casting conditional approval ballots.\\
\textbf{Output:} & A boolean assignment $s^* = (s_1^*,\dots, s_m^*)$ to all issues that achieves $\min_{s\in D}\sum_{i \in [n]} \delta_i(s)$.\\
		\bottomrule
	\end{tabular}
\end{table}

For Example \ref{example:ex1}, we can see that the Conditional Minisum solution would prescribe to the authors to work more for new results, to split their work into two submissions, and not to invite a new co-author, which corresponds to the outcome $\{wm\overline{c}\}$.

If the global dependency graph of an instance is empty, i.e., $\Delta_i=0$ for every voter $i$, then the election degenerates to Unconditional Minisum which is simply the classic Minisum rule in approval voting over multiple independent issues.

Finally, in the following sections, we will extensively make use of the \textit{treewidth of a graph} $G$, denoted as $tw(G)$ \cite{RS86}.
The treewidth of graphs is defined in terms of tree decompositions. A tree decomposition of a graph $G=(N,E)$ consists of a collection of $t$ subsets of $N$, called bags, 
$\{X_1, X_2, \dots, X_t\}$, and a tree $T$, whose node set is the set of bags, such that, $\cup_{i\in [t]}X_i=N,$, for every edge of $E$ there is a bag in $\{X_i: i\in [t]\}$ containing both its endpoints, and for every vertex of $N$ the set of bags containing it induces a connected subtree of $T$. The width of a tree decomposition is the size of the largest bag minus one, i.e.,  $\max_{i\in [t]} |X_i| -1$, and the treewidth of a graph $G$ is the minimum width among all possible tree decompositions of $G$.

\section{Winner Determination}
\label{sec:winner}

In this section we focus on the winner determination problem, under the Conditional Minisum voting rule. In Section \ref{sec:opt}, we mainly present a characterization (subject to certain complexity theory assumptions) of the instances which admit polynomial time algorithms for computing the optimal outcome of the rule, whereas in Section \ref{sec:approx}, our focus is on approximate solutions, for computationally more demanding instances.

\subsection{Optimal Algorithms}
\label{sec:opt}

The price we pay for the higher expressiveness of \cms, compared to the classical Minisum solution, is its increased complexity. Here, we focus on understanding the properties that allow \cms\ to be implemented in polynomial time. For this, we first stick to the case where $\Delta_i\leq 1$ for every voter $i$, which is already \np-hard, and at the same time forms the most obvious, first-step generalization of Unconditional Minisum to the setting of dependencies. Then, in Section \ref{sec:general}, we generalize our results for profiles of bounded $\Delta_i$, for every voter $i$.
To investigate what further restrictions can make the problem tractable, we utilize the global dependency graph of an instance, defined in Section \ref{sec:prelims}, as the aggregation of all the dependencies of the voters into a single graph. 
To see how to exploit the global dependency graph, it is instructive to inspect the \np-hardness proof for \cms\ from the work of \citeauthor{BL16} \citeyear{BL16}, which holds for instances where $\Delta_i = 1$ for every voter $i$, and each dependency graph is acyclic. Examining the profiles created in that reduction, we notice that no restrictions can be stated for the form of the global dependency graph corresponding to the produced instances. This holds since, an acyclic dependency graph for every voter does not necessarily lead to an acyclic global dependency graph and furthermore, the bounded in-degree in each $G_i$, does not imply a constant upper bound for the maximum in-degree of the global graph. 

Our insight is that it may not be only the structure of each voter's dependency graph that causes the problem's hardness, but in addition, the absence of any structural property on the global dependency graph. Motivated by this, we investigate conditions for the global dependency graph, that enable us to obtain the optimal solution in polynomial time. Our findings reveal that this is indeed feasible for the classes of graphs with constant treewidth. For a justification of using the treewidth parameter in the context of elections with interdependent issues, and its practical relevance to the examined setting, we direct the reader to the end of the current subsection.

In our results, we make extensive use of {\em Constraint Satisfaction Problems} (\csps).
A \csp\ instance is described by a tuple $(V, D, C)$, where $V$ is the set of variables, $D$ is the Cartesian product of the domains of the variables, and $C$ is a set of constraints, where a constraint corresponds to a subset of the variables together with all the combinations of values of those variables that make it satisfied. We will pay particular attention to the so-called {\it binary} \csps, where each constraint involves at most two variables. The decision problem for a \csp\ asks whether we can find an assignment to the variables of $V$ so that all constraints of $C$ are satisfied, whereas a natural optimization version \cite{FW92} is to minimize the number of unsatisfied constraints. When analyzing \csps, a useful concept in the literature is the {\it primal} or {\it Gaifman} graph of an instance, defined below.
\begin{definition}
The primal (or Gaifman) graph of a \csp\ instance is an undirected graph, whose vertices are the variables of the instance and there is an edge between two vertices, if and only if they co-appear in at least one constraint.
\end{definition} 

The proof of the following theorem is based on formulating our problem as minimizing the number of unsatisfied constraints in an appropriate binary \csp\ instance, whose primal graph has constant treewidth. For these classes of \csps, one can then use known results by \citeauthor{Freuder90} \citeyear{Freuder90} or by \citeauthor{KHK02} \citeyear{KHK02} for solving them efficiently. 

\begin{theorem}
\label{thm:tw}
If the global dependency graph of a \cms\ instance with $\Delta_i\leq1,$ for every voter $i$, has constant treewidth, then \cms\ can be implemented in polynomial time, even with an arbitrary domain cardinality for each issue.
\end{theorem}

\begin{proof}
Consider an instance $P=(I, D, V, B)$ of \cms\ with $n$ voters and $m$ issues, and let $G$ be its global dependency graph. 
Suppose the treewidth of $G$ is bounded by $k \in \bigoh{(1)}$, and let $d$ be the maximum cardinality among the domains. 
We form an instance of the minimization version of binary \csp, with $m\cdot n$ constraints, where each constraint expresses the satisfaction of a specific voter for a specific issue. 

Recall that we have assumed the maximum in-degree of every voter's dependency graph is at most one, thus each constraint in the \csp\ instance that we construct involves at most two variables, which means that the obtained \csp\ is indeed binary. Also, we can express each constraint by providing at most $d^2$ combinations of the two involved variables (i.e., the combinations that satisfy the constraint). Hence, the construction of the \csp\ instance can be done in polynomial time. 

Since each constraint that involves two variables\footnote{For uniformity, we could add dummy issues in the \cms\ instance (resp. dummy variables in the \csp\ instance) so that the final \csp\ only has constraints with exactly two variables.} corresponds to an edge of the global dependency graph and constraints with exactly one variable do not contribute any edges neither to the primal nor to the global dependency graph, the following can be easily verified. 
\begin{claim}
\label{cl:primal-global-v1}
The primal graph of the produced \csp\ instance is identical to the global dependency graph of the \cms\ instance.
\end{claim}

Therefore, \cms\ has been formulated as minimizing the number of unsatisfied constraints in a binary \csp\ with primal graph of constant treewidth and these classes of CSPs are solvable in $\bigoh{(n^k)}$, \cite{Freuder90,KHK02}.\footnote{In fact, the original results from \citeauthor{Freuder90} \citeyear{Freuder90} do not deal with the optimization version, but as demonstrated in later works (see e.g., Proposition 4.3 from the work of \citeauthor{KKMT19} \citeyear{KKMT19}, it can be extended for this version as well.} 
\end{proof}

We additionally highlight that the above theorem can be generalized when there is a weight $w_i$ for each voter $i$ so that the objective becomes the weighted sum of the dissatisfaction scores. This is simply because the construction from the proof of \Cref{thm:tw} can still be used towards creating an instance of binary (weighted) \csp, where all constraints derived from the preferences of a voter $v_i$ are weighted by $w_i$, the weight of $v_i$. Such an instance can again be solved using the same algorithm that was applied to the original (unweighted) \csp\ instance since in the work of \citeauthor{KKMT19} \citeyear{KKMT19} the definition of the \csp\ problem explicitly includes weighted constraints.

In trying to move away from treewidth-based assumptions, a natural question is whether we can solve other classes of instances, containing graphs of non-constant treewidth, by focusing on other parameters of the problem. Quite surprisingly, it turns out that bounded treewidth is essentially the only property that can yield efficiency guarantees. To establish this claim, we will first show a ``reverse" direction to Theorem \ref{thm:tw}, namely that binary \csps\ can be reduced to solving \cms. Hence, together with Theorem \ref{thm:tw}, this means that \cms\ is computationally equivalent to binary \csps, and thus to any other problem for which the same result has been already established, e.g., the \textsc{partitioned subgraph isomorphism} problem \cite{M10}.

\begin{theorem}
\label{thm:tw-hardness}
Every binary CSP with primal graph $G$, can be reduced in polynomial time to a \cms\ instance with $\Delta_i\leq 1$ for every voter $i$, and with $G$ as the global dependency graph.
\end{theorem}

\begin{proof}
For convenience, we will work with the standard decision version of \csp\ where one asks if there is a solution that satisfies all the constraints. 

Let $P$ be a binary \csp\ instance, and without loss of generality, assume that every constraint involves exactly two variables (which can be enforced by the addition of dummy variables). We construct a \cms\ instance $P'$, where the issues correspond to the variables and the voters correspond to the constraints of $P$. In particular, for every variable $x_j$ of the \csp\ instance, we add an issue $I_j$ and for every constraint we add a voter, with the following preferences: let $x_j, x_k$ be the two variables involved in the constraint. We pick one of the two variables (arbitrarily), say $x_k$, and we set $I_k$ as the issue that the voter cares about, conditioned on $I_j$. We also set her conditional ballot for issue $I_k$ in such a way, so that the voter becomes satisfied precisely for all combinations of values for $x_j$ and $x_k$ that make the constraint satisfied. The voter is also satisfied unconditionally with every outcome for every other issue of the produced instance. Obviously, the dependency graph of every voter has maximum in-degree equal to one.

As an example, suppose that a constraint is of the form $x_1 \vee x_2$ and the variables $x_1,x_2$ have binary domain. Then we introduce a new voter, and two issues, $I_1$, $I_2$ (the issues may have been introduced already by other constraints in the instance), and we can select $I_2$ as being dependent on $I_1$. The conditional ballot regarding the satisfaction of the voter for $I_2$ is $\{x_1:x_2\}$, $\{\overline{x_1}:x_2\}$, $\{x_1:\overline{x_2}\}$. In addition, the voter has an unconditional ballot for $I_1$, in the form $\{x_1,\overline{x_1}\}$, thus approving every value for $I_1$. 

To complete the reduction, we consider the decision version of \cms\, where we ask if there is an assignment with no dissatisfactions, i.e., the instance $P'$ has an affirmative solution only when all voters are satisfied with all the issues. It is obvious that this is a polynomial time reduction 
(the conditional ballot of each voter for her single issue of interest can be described in $\bigoh{(d^2)}$ time, where $d$ is the maximum domain cardinality of the \csp\ variables). 
It is quite obvious also that every edge from the primal graph of $P$ corresponds to an edge in the global dependency graph of $P'$, and vice versa. Hence:
\begin{claim}
\label{cl:primal-global}
The primal graph of \csp\ instance $P$ is identical to the global dependency graph of the \cms\ instance $P'$.
\end{claim}

Finally, it remains to see that there exists a solution to $P'$ if and only if there exists a solution to $P$. Indeed, any solution to $P'$ corresponds to an assignment of values to the issues such that all voters are satisfied with all issues, which means that all the constraints of the \csp\ instance $P$ are satisfied. The converse is also easily verified.
\end{proof}

Theorem \ref{thm:tw-hardness} allows us to apply some well known hardness results on binary \csps\ \cite{Grohe07,GSS01}, which imply that one cannot hope to have an efficient algorithm for a class of \cms\ instances, if the class contains instances with non-constant treewidth.
Hence, Theorem \ref{thm:tw} is essentially tight, and this resolves the problem of finding a characterization for polynomial time solvability of \cms,
subject to a standard complexity theory assumption. This is summarized in the following corollary.

\begin{corollary}[implied by Theorems \ref{thm:tw} and \ref{thm:tw-hardness} and by \citeauthor{Grohe07} \citeyear{Grohe07}]
\label{corol:tw-hard}
    Let ${\cal G}$ be a recursively enumerable (e.g., decidable) class of graphs, and let $\cms({\cal G})$ be the class of instances with a global dependency graph that belongs to ${\cal G}$, and with $\Delta_i\leq 1$ for every voter $i$. Assuming  FPT $\neq$ W[1], there is a polynomial algorithm for $\cms({\cal G})$ if and only if every graph in ${\cal G}$ has constant treewidth modulo homomorphic equivalence.
\end{corollary}
\begin{proof}
If ${\cal G}$ is a class of graphs, as in the statement, then by Theorem \ref{thm:tw-hardness}, an algorithm for the class of \cms\ instances whose global dependency graph belongs to ${\cal G}$ implies an algorithm for the \csp\ instances whose primal graph belongs to ${\cal G}$. The proof can now be completed by applying the hardness results for binary \csps\ \cite{Grohe07,GSS01}.
\end{proof}

The characterization provided in \Cref{corol:tw-hard} is unexpectedly strong. It follows from a well-known and fundamental result from the literature on \csps\ \cite{Grohe07}. We note that \Cref{corol:tw-hard} is not a result about parameterized complexity, but rather a statement about polynomial-time solvability, grounded in a complexity-theoretic assumption concerning classes frequently used in the parameterized complexity literature. Quoting \citeauthor{Grohe07} \citeyear{Grohe07}: “This [the usage of the FPT $\neq$ W[1] assumption] is remarkable, because the statement of the theorem has nothing to do with parameterized complexity theory.”
Regarding the parameterized complexity of the examined problem, recall that the proof of \Cref{thm:tw} implies membership in XP when parameterizing by the treewidth parameter, and we refer to \Cref{sec:param} for further results on these aspects. The hypothesis that $\mathcal{G}$ is recursively enumerable can be dropped by slightly strengthening the computational complexity assumption; we refer to the work of \citeauthor{Grohe07} \citeyear{Grohe07} for more details.

\begin{remark}
\label{rem:tw}
If we strengthen the complexity assumption used in Corollary \ref{corol:tw-hard}, from FPT $\neq$ W[1] to the Exponential Time Hypothesis (ETH), we can obtain an even stronger impossibility. In particular, by exploiting the result of \citeauthor{M10} \citeyear{M10}, and the proof of Theorem \ref{thm:tw-hardness}, we can show that under ETH, one cannot even hope for an algorithm on $\cms({\cal G})$ that runs in time $f(G) ||P||^{o(tw(G))/log(tw(G))}$, where $||P||$ is the size of the \cms\ instance and $G\in {\cal G}$. This implies that the running time $\bigoh{(n^{tw(G)})}$ of the algorithm from Theorem \ref{thm:tw} is the best possible up to an $\bigoh{(\log{(tw(G))})}$ factor in the exponent.
\end{remark}

\subsubsection{Generalizations to higher in-degrees}
\label{sec:general}

We highlight that Theorem \ref{thm:tw} cannot be immediately generalized so as to apply to instances where $\Delta_i\geq 2$ for some voter $i$, since in that case the global dependency graph will not necessarily coincide with the primal graph of the corresponding \csp\ that we constructed in the proof of Theorem \ref{thm:tw} (which is an essential part of the proof). In order to obtain a result for higher in-degrees, we introduce the following definition, which is a generalization of the global dependency graph, and where we simply replace a vertex and its in-neighbors by a clique on the same set of vertices.

\begin{definition}
The extended global dependency graph of a set of voters is the undirected (simple) graph $(I,\bigcup_{i\in[n]}\tilde{E_i})$, where $\tilde{E_i}$ is the edge set of a graph that corresponds to voter $i$ and is created by enforcing an undirected clique for every issue $I_j$ and any voter $i$, on the set $N^{\unaryminus}_{i}(I_j) \cup \{I_j\}$. 
\end{definition}

Note that for the cases that $\Delta\leq1$, the extended global dependency graph of an instance coincides with the global dependency graph (and hence with the primal graph created in the proof of Theorem \ref{thm:tw}). The crucial observation now is that 
as long as $\Delta\in O(1)$, an instance of \csp\, equivalent to the initial \cms\ instance, can be created in polynomial time by following very closely the proof of Theorem \ref{thm:tw}. And most importantly, even though the extended global dependency graph of the \cms\ instance does not coincide with the global dependency graph, it does coincide with the primal graph of the created \csp\ instance, which is all we need. We stress also that one of the reasons we need $\Delta\in O(1)$, is to ensure that we can list all the combinations that satisfy a constraint in polynomial space, namely to ensure that the combinations of values of variables that satisfy the constraint, which are  are at most $d^\Delta$, are no more than polynomially many. 

To finalize the argument for the generalization, note that the created \csp\ instance will no longer be a binary \csp\ (i.e., it will not have at most two variables in each constraint). Nevertheless, these instances will have at most a constant number of variables in each constraint, due to $\Delta$ being constant, and they are still tractable as long as the primal graph of the \csp\ has bounded treewidth \cite{Freuder90}. Hence, our discussion can be summarized by the following theorem, which is indeed a generalization of Theorem \ref{thm:tw} for instances of higher in-degrees.

\begin{theorem}
\label{thm:twgen}
If the extended global dependency graph of a \cms\ instance with $\Delta_i \in O(1)$ for every voter $i$, has constant treewidth, then \cms\ can be implemented in polynomial time, even with an arbitrary domain cardinality for each issue.
\end{theorem}

Finally, we can also obtain a generalization of Theorem \ref{thm:tw-hardness} (the exact same arguments apply with the global dependency graph being replaced by the extended global dependency graph). This leads to the following characterization regarding instances with higher in-degrees, which is the analog of Corollary \ref{corol:tw-hard}. 

\begin{corollary}
\label{corol:tw-hard-general}
    Let ${\cal G}$ be a recursively enumerable class of graphs, and let $\cms({\cal G})$ be the class of instances with an extended global dependency graph that belongs to ${\cal G}$, and with $\Delta_i\in O(1)$ for every voter $i$. Assuming  FPT $\neq$ W[1], there is a polynomial algorithm for $\cms({\cal G})$ if and only if every graph in ${\cal G}$ has constant treewidth.
\end{corollary}

We finally note that Remark \ref{rem:tw} also applies here, but again for the treewidth of the extended global dependency graph.

\subsubsection{Parameterized complexity of {\textsc{cms}}} 
\label{sec:param}

The algorithm used in the proof of Theorem \ref{thm:tw}, runs in time exponential in $tw(G)$, where $G$ is the global dependency graph and thus it places \cms\ in XP w.r.t the treewidth parameter. One can wonder if anything more can be said concerning the fixed parameter tractability of the problem. Given the equivalence of our problem for $\Delta_i\leq 1$, for every voter $i$, with binary \csp, we can use existing results \cite{SS10,GS08} to extract some further characterizations and obtain an almost complete picture with respect to the most relevant parameters. On the positive side, we can see that our problem is in FPT with respect to the parameter ``treewidth + domain size". On the negative side, we cannot hope to prove FPT only w.r.t the one of the two parameters, independent of the other, as stated below.

\begin{corollary}
\label{corol:param}
When $\Delta_i\in \bigoh{(1)}$ for every voter $i$, \cms\ is in FPT w.r.t the parameter $tw+d$, where $tw$ is the treewidth of the extended global dependency graph and $d$ is the maximum domain size. Moreover, even when $\Delta_i\leq 1$ for every voter $i$, it is $W[1]$-hard with respect to $tw$ and with respect to $d$.
\end{corollary}

\begin{proof}
First, let us introduce some notation for ease of presentation, following the conventions from \cite{SS10}. 
Given a set of parameters $S$, we denote as $\Pi\{S\}$ the parameterized version of a decision problem $\Pi$, having all variables in $S$ as parameters; formally this corresponds to a single parameter whose value is upper bounded by an integer $\sigma$, defined as the sum of all parameter values in $S$, and we say that $\Pi\{S\}$ is in FPT when $\Pi$ is in FPT with respect to $\sigma$.
For a \csp\ instance we will denote by $tw'$ the treewidth of its primal graph, by $d'$ the maximum domain size of every variable, and by $arity$ the maximum number of variables that co-appear in a constraint. 

To prove the positive statement, we exploit the fact that $\csp\{arity,d',tw'\}$ is in FPT \cite{SS10,GSS01}. This trivially implies that for \csp\ instances of constant arity, we have that $\csp\{d',tw'\}$ is in FPT. We can now use our Theorem \ref{thm:twgen}. In particular, if we have a \cms\ instance, where $\Delta_i\in \bigoh{(1)}$, and where $d$ is the maximum domain size and $tw$ is the treewidth of the extended global dependency graph, Theorem \ref{thm:twgen} shows that we can reduce this to solving a \csp\ instance of constant arity and with $d'=d$ and $tw'=tw$. Hence, we have that $\cms\{tw, d\}$ is in FPT, when $\Delta_i\in \bigoh{(1)}$.

To prove the negative statements, we use the following definition: A set of parameters $S$ dominates a set $S'$ if whenever all parameters of $S'$ are bounded by some constants, all parameters of $S$ are bounded too. In the work of \citeauthor{SS10} \citeyear{SS10} (Theorem 1 therein), it was proved that $\csp_{bin}\{arity, tw'\}$ and $\csp_{bin}\{arity, d'\}$ are W[1]-hard, where $\csp_{bin}$ denotes the class of binary \csp\ instances. It is trivial to see that the set $S = \{tw'\}$ dominates the set $S' = \{arity, tw'\}$. Hence, by utilizing Lemma 1 from the work of \citeauthor{SS10} \citeyear{SS10}, we obtain that $\csp_{bin}\{tw'\}$ is $W[1]$-hard and the same is true also for $\csp_{bin}\{d'\}$. Given the reduction established in our Theorem \ref{thm:tw-hardness}, of binary \csps\ with parameters $tw'$ and $d'$ to \cms\ instances with $\Delta_i \leq 1$ and with $tw=tw'$ and $d=d'$, we can conclude that  
both $\cms\{tw\}$ and $\cms\{d\}$ are W[1]-hard too.
\end{proof}

We conclude the subsection by noting that the instances captured by the assumptions we have made are indeed meaningful in multi-issue elections with logically dependent issues. First, we mostly considered instances where $\Delta_i\leq 1$ for every voter $i$, which is non-trivial (\np-hard to solve the winner determination problem), but still the obvious first-step generalization of the traditional (minisum) approval voting rule. Secondly, the main positive result was for the case where the global dependency graph has a bounded treewidth. This allows for various types of graphs, such as disjoint edges representing pairs of mutually correlated issues, or paths (and cycles) which can represent issues that are sequentially dependent. When the global dependency graph forms a tree, it creates a hierarchical structure of dependencies. For example, out-stars occur when central issues influence the satisfaction of voters on other issues, while in-stars represent cases where multiple issues affect central ones. Going further, a constant treewidth allows for even more complex dependencies among issues, but still well-structured.


\subsection{Approximation Algorithms}
\label{sec:approx}

It is well known that a Minisum solution can be efficiently computed when there are no dependencies \cite{BKS07}. In contrast to this, \cms\ is \np-hard even when all the issues have a binary domain and there is only a single dependence per voter, i.e., when every voter's dependency graph has just a single edge \cite{BL16}. Given this hardness result, it is natural to resort to the framework of approximation algorithms. The only known result from this perspective is an algorithm by \citeauthor{BL16} \citeyear{BL16}, with a {\emph {differential}} approximation ratio of $4.34/(m\sum_{j\in{I}} 2^{|N^{\unaryminus}(j)|}+4.34)$, for the case of a common acyclic dependency graph, where $N^{\unaryminus}(j)$ is the set of common in-neighbors of issue $j$ (for each voter $i$ and issue $j$, $N^{\unaryminus}_i(j) = N^{\unaryminus}(j)$). However, differential approximations (we refer to the work of \citeauthor{DGP98} \citeyear{DGP98} for the definition of this concept) form a less typical approach in the field of approximation algorithms. 
Instead, we focus on the more standard framework of {\it multiplicative} approximation algorithms, as treated also in common textbooks \cite{Vazirani03,WS11}.
We say that an algorithm for a minimization problem achieves a multiplicative ratio of $\alpha \geq 1$, if for every instance $I$, it produces a solution with cost at most $\alpha$ times the optimal.
We stress that a differential approximation ratio for minimization problems does not in general imply any multiplicative approximation ratio \cite{BP03}.

We start first with a rather strong negative result in terms of the viability of approximate solutions. The main result of the previous subsection was the hardness of computing optimal outcomes, which is implied by Theorem \ref{thm:tw-hardness}. 
In fact, the proof of Theorem \ref{thm:tw-hardness} implies also the following important multiplicative inapproximability about \cms.

\begin{corollary}
\label{cor:no-approx}
Even when $\Delta_i\leq 1$ for every voter $i$, it is \np-hard to obtain any finite approximation ratio for \cms.
\end{corollary}

\begin{proof}
If we look again at the proof of Theorem \ref{thm:tw-hardness}, we can see that we have reduced the solution of a binary \csp\ instance to deciding whether a \cms\ instance admits a solution of cost zero, i.e., a solution where all voters are satisfied. Given the hardness of binary \csps, we conclude then that deciding if a \cms\ instance has optimal cost equal to zero is \np-hard. Suppose now that we could obtain an approximation algorithm with some finite approximation ratio for every instance. This immediately means that we could use this algorithm to distinguish between instances that have an optimal cost of zero (where the algorithm would have to return the optimal solution by the definition of approximation ratio) from the remaining instances (where the algorithm would return some solution with a positive cost). Hence we would have solved an \np-hard problem.
\end{proof}

Therefore, a polynomial time algorithm with a bounded multiplicative approximation guarantee, could only be possible under further assumptions. 
Our main contribution in this subsection is the first class of multiplicative approximation algorithms for some special cases of \cms. 
Sticking to the already hard class of binary domains and $\Delta_i\leq 1,$ for every voter $i$ (which includes the instances in which every voter has one edge, considered in the hardness result by \citeauthor{BL16} \citeyear{BL16}), we focus on instances that satisfy an assumption motivated by the fact that in the unconditional case, allowing voters to approve at most a single alternative per issue is already an interesting and well-studied voting scenario, which corresponds to the multi-issue analog of elections under the classical plurality setting.

\begin{definition}
    Consider a \cms\ instance with binary domains, and where the dependency graph of every voter $i$ satisfies $\Delta_i\leq 1$. The instance is called $1$-approval, if for every issue $I_j$ that is dependent on some issue $I_k$ according to the preferences of a voter $i$, it holds that $i$ can be satisfied only with one pair, say $\{x_k:x_j\}$, w.r.t. $I_j$, where $x_k\in D_k$ and $x_j\in D_j$. No restrictions are imposed to the number of approved alternatives for unconditional ballots.
\end{definition}

To obtain a positive result, we first make use of known approximation algorithms for \textsc{min $k$-sat}, a minimization version of \textsc{sat}, where we are given a set of $m$ clauses in $k$-CNF and we search for a boolean assignment so as to minimize the total number of satisfied clauses. 
Interestingly, minimization versions of \textsc{sat} have rarely been applied in the context of computational social choice.
In fact it has hardly ever been used as a tool for obtaining approximation algorithms for other problems (we are only aware of an application for certain string comparison problems \cite{GKZ05}). The use of \textsc{max sat} is much more common, but for the case of \cms, and for multiplicative approximation guarantees, it does not seem convenient to exploit algorithms for maximisation problems.
In a nutshell, if we use an approximation algorithm for \textsc{max sat}, the conversion from the solution of a maximization problem to that of a minimization one that we have here, does not preserve a good approximation ratio for our objective function\footnote{The differential approximation result \cite{BL16} was based on the use of \textsc{max sat}. But as stated earlier, this does not imply any non-trivial multiplicative approximation for \cms.}.
The main positive result of this subsection follows.

\begin{theorem}
	\label{thm:approx-msat}
Let $\mathcal{F}$ be the family of $1$-approval \cms\ instances, with binary domains and with $\Delta_i\leq 1$ for every voter $i$. Then any $\alpha$-approximation algorithm for \textsc{min 2-sat} yields an $\alpha$-approximation for any instance in $\mathcal{F}$. In particular, we can have a polynomial time $1.1037$-approximation for any instance in $\mathcal{F}$. 
\end{theorem}

\begin{proof}
We present a reduction to \textsc{min 2-sat} that preserves the approximation factor in the case where the given \cms\ instance is 1-approval. We first present a general reduction for any instance with $\Delta_i\leq 1$ for every voter $i$, which could be of broader interest. Later on, we will see how we can exploit this construction for $1$-approval instances. 
Therefore, consider an arbitrary instance $P$ of $n$ voters, with $\Delta_i\leq 1$ for every voter $i$, and let $I = \{I_1, \dots, I_m\}$ be the set of issues. We first create a logical formula $C_{ij}$, for every voter $i\in V$, and every issue $I_j\in I$, which indicates the cases where voter $i$ is {\it{not}} satisfied with the outcome on $I_j$. For every issue $I_j$, recall that $D_j=\{d_j,\overline{d_j}\}$ is its domain, and $x_j$ will be the corresponding boolean variable in the construction of $C_{ij}$.

	For this we consider two cases. The first and easier case is when for a voter $i$, and issue $I_j$, $N^{\unaryminus}_{i}(I_j)=\emptyset$. All possible forms of $B_i^j$ are depicted in the first row of Table \ref{tab:no-pred}, whereas the corresponding formula is shown in the second row.

	\begin{table}[H]
	\centering
\begin{tabular}{c|c|c|c|c}
$B_i^j$ & $\emptyset$ &  $\{d_j\}$   & $\{\overline{d_j}\}$  &  $\{d_j,\overline{d_j}\}$ \\ \hline       
$C_{ij}$ & $x_j\vee\overline{x}_j$ & $\overline{x}_j$ & $x_j$ & $\emptyset$
	\end{tabular}
	\caption{The formula when issue $I_j$ has no predecessor in $G_i$. \label{tab:no-pred}}
\end{table}
	
On the other hand, if $I_j$ has an in-neighbor (it can have only one by our assumptions), say $I_k\in I$, we set $C_{ij}$ equal to the disjunction of all combinations of outcomes on issues $I_j$ and 
$I_k$ that dissatisfy voter $i$ with respect to $I_j$. 
To illustrate this construction, we describe an example with $4$ voters, $2$ issues $I=\{I_k,I_j\}$ and for every voter $i,$ $G_i=\{I,\{I_k,I_j\}\}$. The preferences for issue $I_j$ are shown in Table \ref{tab:pred}. Namely, for $i=1,2,3,4$, the first cell in the $i$-th row depicts $B_i^j$ from which $C_{ij}$ can be obtained as the disjunction of the ticked expressions in the remaining of the $i$-th row.

\begin{table}[htbp]
\centering
\begin{tabular}{ C{2.7cm}|C{ 0.95cm}C{ 0.95cm}C{ 0.95cm}C{ 0.95cm}  }
\diagbox{\hspace{0.8cm}{\scalebox{.8}{$B_i^j$}}}{\hspace{0.8cm} }  & {\scalebox{.8}{ $(x_k\wedge x_j)$ }}& {\scalebox{.8}{$(x_k \wedge \overline{x}_j )$}} &{\scalebox{.8}{ $( \overline{x}_k \wedge x_j )$}} &{\scalebox{.8}{ $( \overline{x}_k \wedge \overline{x}_j )$}} \\
\hline
$\emptyset$                                         &         \checkmark                               & \checkmark     & \checkmark     & \checkmark      \\ \hline

$\{d_k:d_j\}$                                        &                                &       $\checkmark$       & \checkmark     & \checkmark      \\\hline

 \begin{tabular}{@{}c@{}}$\{d_k:d_j\}$,\\ $\{d_k:\overline{d_j}\}$\end{tabular}                             &                                        &      &         \checkmark   & \checkmark      \\\hline

 \begin{tabular}{@{}c@{}}$\{d_k:d_j\},$ \\ $\{\overline{d_k}:d_j\}\!,\!  \{d_k:\overline{d_j}\}$\end{tabular}
 &                    &            &            & \checkmark      
       
\end{tabular}
	\caption{For $i=1,2,3,4$ the formula $C_{ij}$ is the disjunction of the ticked expressions in the $i$-th row.} \label{tab:pred}
\end{table}

\begin{claim}
For an outcome $(s_1,\dots, s_m)$ of the issues and the corresponding assignment to the boolean variables $x_1,\dots, x_m$, voter $i$ is dissatisfied with $I_j$ if and only if the formula $C_{ij}$ is true.
\end{claim}

The constructed formula $C_{ij}$ is in DNF. To continue, we will need to make a conversion to CNF, which is easy to do given its small size as per the following lemma.

\begin{lemma}
	\label{lem:DNFtoCNF}
		The formula $C_{ij}$ for each voter $i\in V$, and each issue $I_j\in I$, can be written in CNF with at most 2 clauses, and where each clause contains at most 2 literals.
	\end{lemma}

\begin{proof}[Proof of Lemma \ref{lem:DNFtoCNF}]
Fix a voter $i$ and an issue $I_j$. For the cases where issue $I_j$ has no in-neighbor in $G_i$, the lemma obviously holds, as can be verified in Table \ref{tab:no-pred}. For all other cases, $I_j$ has a unique in-neighbor, say issue $I_k$, since we are dealing with instances where the in-degree is at most one. We now need to examine the form of $C_{ij}$ for the cases that arise. 
\begin{description} 
\item[Case A.] 	
If voter $i$ is satisfied only with $1$ out of the $4$ possible outcomes regarding $I_j$ and $I_k$, then $C_{ij}$ is a disjunction of $3$ conjunctions. Let us assume that $C_{ij}$ is in the form: 
	$(x_j\wedge x_k)\vee(\overline{x}_j\wedge x_k)\vee (x_j\wedge \overline{x}_k)$. All other cases are handled in exactly the same way. The following equivalences can bring $C_{ij}$ to the desirable form.
	\begin{align*}
		(x_j\wedge x_k)\vee(\overline{x}_j\wedge x_k)\vee (x_j\wedge \overline{x}_k)&\equiv x_k\vee (x_j\wedge \overline{x}_k) \equiv\\
		 (x_k \vee x_j) \wedge (x_k \vee \overline{x}_k) &\equiv x_k \vee x_j
	\end{align*}
\item[Case B.] If voter $i$ is satisfied with $2$ out of the $4$ possible outcomes, then $C_{ij}$ is a disjunction of $2$ conjunctions. Without loss of generality, we can assume we have one of the following cases (all remaining cases can also be brought to one of these formats). As verified below, by the right hand side of each term, all cases can be brought into the desirable form. 
	\begin{align*}
(1.) \hspace{0.7cm} (x_j\wedge \overline{x}_k)\vee(\overline{x}_j \wedge x_k) &\equiv (x_j \vee x_k)\wedge (\overline{x}_j \vee \overline{x}_k)\\
(2.) \hspace{0.7cm}(x_j\wedge x_k)\vee(\overline{x}_j \wedge \overline{x}_k) &\equiv (x_j \vee \overline{x}_k)\wedge (\overline{x}_j \vee x_k)
	\end{align*}
 \item[Case C.] If voter $i$ is satisfied with 3 out of the $4$ possible outcomes regarding $I_j$ and $I_k$, then we simply take the conjunction expressing the outcome that causes dissatisfaction. E.g., $C_{ij} = x_k\wedge \overline{x}_j$, when $i$ is satisfied with everything apart from $\{d_k:\overline{d_j}\}$. Thus, $C_{ij}$ has 2 clauses with 1 literal each.\end{description}

Note that, typically, according to the definition of the conditional approval framework, a voter could also be satisfied with all $4$ possible outcomes of $I_j$ and $I_k$ or disagree with all possible outcomes.
However, one can consider such ballots as simple unconditional ballots, as no real dependence between the issues exists.  
\end{proof}

Using Lemma \ref{lem:DNFtoCNF} to convert each $C_{ij}$ to CNF, we can now create a \textsc{min 2-sat} instance $P'$ by the multiset\footnote{Some clauses may happen to appear more than once in the final formula but there is no harm in keeping such duplicates.} of all clauses appearing in the $C_{ij}$'s, i.e., appearing in the formula
\begin{equation}
	\label{msat_instance}
C= \bigwedge_{i\in V, I_j \in I}C_{ij}.
\end{equation}

We now try to exploit how the analysis so far can help us for $1$-approval instances. We will first need to compare the optimal solution of the \cms\ instance with the optimal solution of the corresponding \msat\ instance.

	\begin{lemma} 
	\label{lem:opt} Let $P$ be a \cms\ instance and $P'$ be its corresponding \msat\ instance produced as discussed in the proof of Theorem \ref{thm:approx-msat}. Let also $\opt(P)$ and $\opt(P')$ be the values of the optimal solutions of the instances. If $P$ is $1$-approval, then it holds that $\opt(P')= \opt(P)$.
	\end{lemma}

\begin{proof}[Proof of Lemma \ref{lem:opt}]	
Consider an optimal solution of the \cms\ instance $P$. Every voter contributes to the cost of this solution precisely the number of issues with which she is dissatisfied. Consider now the corresponding \msat\ instance $P'$, formed by the clauses of the constructed formula $C$ from Equation \eqref{msat_instance}. Let us look at the truth assignment to the variables of $C$, as dictated by the values of the issues in the optimal solution of $P$. We will provide an upper bound on the number of satisfied clauses of $C$. 
Under this truth assignment, it holds that for every voter $i$ and for every issue $I_j$ for which $i$ is dissatisfied with respect to $I_j$, the formula $C_{ij}$ is true. By Lemma \ref{lem:DNFtoCNF}, any $C_{ij}$ has at most two clauses which could be satisfied when $C_{ij}$ is true. But in the case when $P$ is $1$-approval, then any fixed voter $i$ either votes unconditionally on $I_j$ or her ballot belongs to Case A from the proof of Lemma \ref{lem:DNFtoCNF}. In both cases, $C_{ij}$ is formed by a single clause. Hence, by looking at all the clauses of $C$ that come from combinations $(i, j)$, for which voter $i$ is dissatisfied with respect to issue $I_j$, we get a number of satisfied clauses equal to $\opt(P)$. Let us focus now on pairs $(i, j)$, for which voter $i$ is satisfied with respect to $I_j$. Then, the corresponding formula $C_{ij}$ is false. If the ballot of voter $i$ with respect to $I_j$ is unconditional or if her ballot corresponds to Case A of Lemma \ref{lem:DNFtoCNF}, then $C_{ij}$ does not have any true clauses.
Therefore, for $1$-approval instances, $\opt(P')$, which is the total number of satisfied clauses of $C$ under the selected truth assignment, equals 
$\opt(P)$. 
\end{proof}

Our construction gives rise to the following algorithm for \cms, under the discussed assumptions:

\begin{figure}[H]
  \makebox[\linewidth]{%
  \begin{minipage}{\dimexpr\linewidth-12.2em}
\begin{algorithm}[H]
\caption{\hfill {$\rhd$Input: 1-approval profile P}\label{alg:approx-msat}}
\begin{algorithmic}[1]
\STATE Create $P'$ from $P$ using Lemma \ref{lem:DNFtoCNF} and Equation \eqref{msat_instance}.
\STATE Run an $\alpha$-approximation of \msat\ on $P'$.
\STATE Set the value of $I_j$ in $P$ to the value of $x_j$ in $P'$.
\end{algorithmic}
\end{algorithm} \end{minipage}}
\end{figure}

To conclude the proof of Theorem \ref{thm:approx-msat}, let $\sol(P')$ be the cost of the solution to $P'$ produced in step 2 of Algorithm \ref{alg:approx-msat}, which equals the number of satisfied clauses in $C$ by the truth assignment of the $\alpha$-approximation algorithm. This corresponds to a solution for \cms\ and let $\sol(P)$ be its total cost. We note that the total number of distinct pairs $(i, j)$ for which voter $i$ is dissatisfied by issue $I_j$ can be no more than the number of the satisfied clauses of $C$, since each $C_{ij}$ corresponds to a pair of a voter and an issue.
Hence, together with Lemma \ref{lem:opt}, we have the following implications:
$$ \sol(P) \leq \sol(P') \leq \alpha\cdot \opt(P') = \alpha\cdot \opt(P)$$

Thus, every $\alpha$-approximation algorithm for \textsc{min 2-sat} yields an $\alpha$-approximation for \textsc{cms}, as long as $P$ is $1$-approval. To obtain the claimed approximation ratio, we use the algorithm for \textsc{min 2-sat} \cite{AZ05} achieving a factor of 1.1037. 
\end{proof}

\begin{remark}
The proof of Theorem \ref{thm:approx-msat} also reveals why we cannot extend it to have a constant approximation for other than $1$-approval instances. In particular, for instances that involve the Cases B and C, described in the proof of Lemma \ref{lem:DNFtoCNF}, we cannot guarantee that Lemma \ref{lem:opt} will hold (all we need is that $OPT(P') \leq OPT(P)$, but this could be far from true).     
\end{remark}

Although we have not been able to obtain a constant factor approximation for any other instance of \cms, the proof of Theorem \ref{thm:approx-msat} motivates the study of two more special cases of interest, for which we can obtain a positive result via different procedures. 
The central idea is that the general construction presented in Lemma \ref{lem:DNFtoCNF} identifies 2 more cases, other than $1$-approval, that may occur regarding the satisfaction of a voter w.r.t. an issue. These are precisely the Cases B and C in the proof of Lemma \ref{lem:DNFtoCNF}. 
To define these two cases more formally, once again, we consider instances with binary domains, and with $\Delta_i\leq 1$ for every voter $i$. We will provide positive results for the families of instances that will be called OR-instances and XOR-instances. The former family could be seen as a generalized variant of antiplurality instances (a.k.a. veto, see e.g. \citeauthor{handbook} \citeyear{handbook} for more details); it contains instances in which every voter who casts a ballot for an issue $I_j$ that is conditioned on the outcome of an issue $I_k$, approves exactly three out of the four possible combinations for these issues, or equivalently, is dissatisfied only with a single combination (the reason behind the name of this family will become clear when inspecting the proof of the approximate result that follows). The latter, includes instances in which for every issue $I_j$ that is dependent on an issue $I_k$ according to the preferences of a voter $i$, we assume that these issues are of a complementary nature, i.e., that voter $i$ either wants $I_j$ to be set to the same value as $I_k$ or to the opposite (but not both). In other words, the satisfaction of the voter depends on the XOR value between $I_j$ and $I_k$. In both families, we impose no restrictions for the issues that have no dependence on other issues. The formal definitions of the instances that we are going to examine follow.

\begin{definition}
    We say that a \cms\ instance where the issues are binary and the dependency graph of every voter $i$ satisfies $\Delta_i\leq 1$ is an OR-instance if every voter who casts a ballot on an issue $I_j$ that is conditioned on the outcome of an issue $I_k$ is approving all but one combinations $\{x_k:x_j\}$, where $x_\ell \in \{d_\ell,\overline{d_\ell}\},$ for $\ell\in \{k,j\}$.
\end{definition}

\begin{definition}
    We say that a \cms\ instance where the issues are binary and the dependency graph of every voter $i$ satisfies $\Delta_i\leq 1$ is a XOR-instance if every voter who casts a ballot on an issue $I_j$ that is conditioned on the outcome of an issue $I_k$, is voting either for 
    $\{d_k:d_j,\overline{d_k}:\overline{d_j}\}$ or for $\{d_k:\overline{d_j},\overline{d_k}:d_j\}$.
\end{definition}

In contrast to the proof of Theorem \ref{thm:approx-msat}, we are now going to reduce to and use algorithms for the \textsc{min-2-cnf-deletion} problem for OR-instances and the \textsc{min-uncut} problem for XOR-instances. Likewise in the work of \citeauthor{agarwal2005log} \citeyear{agarwal2005log}, we are going to define these problems in a unified formulation that is convenient to us, and we will consider them as special cases of the \textsc{constraint satisfaction problem (csp)} which also appeared in Section \ref{sec:opt}. Say that we are given a set of boolean variables $b_1,\dots,b_n$ and a set of constraints $C$ and the
goal is to find an assignment that minimizes the number of
unsatisfied constraints. The \textsc{min-2-cnf-deletion} problem is the special case of \textsc{constraint satisfaction problem} in which each constraint can be written in a 2-CNF form. More precisely we will focus on instances in which each constraint corresponds to a single clause in 2-CNF form, which has been called \textsc{2-cnf clause-deletion} problem in the literature \cite{klein1997approximation}. The \textsc{min-uncut} problem is the special case of \textsc{constraint satisfaction problem} in which each constraint is of the form $b_i \xor b_j = 0$ or $b_i \xor b_j = 1$ and has also been called \textsc{2-cnf{\normalfont{$\equiv$}} deletion} in the literature \cite{garg1996approximate}. 

\begin{theorem} \label{approx:non-const}
For every XOR-instance (resp. OR-instance) $P$ it holds that any $\alpha$-approxi\-mation algorithm for \textsc{min-uncut} (resp. \textsc{2-cnf clause-deletion}) yields an $\alpha$-approxi\-mation for \cms\ in $P$. In particular, we can have a polynomial time $\log{m}$-approximation for the class of XOR-instances and a polynomial time $\log{m}\log\log{m}$-approximation for the class of OR-instances.
\end{theorem}

\begin{proof}
    We start by proving the statement for XOR-instances. The result follows from an approximation preserving reduction of an instance $P$ to an instance $P'$ of \textsc{min-uncut}. This reduction is similar to the one presented in the proof of Theorem \ref{thm:approx-msat} but, in this case, the satisfaction of a voter with respect to an issue should correspond to a satisfied constraint. Consider a voter $i$ of $P$ that has an unconditional ballot with respect to an issue $I_j$, say in favor of the alternative $x_j\in D_j$ (resp. $\overline{x_j}\in D_j$), then, her preference can be simply expressed as $x_j\xor 0=1$ (resp. $x_j\xor 1=1$). On the other hand, if the voter's ballot on $I_j$ is conditioned on the outcome of $I_k$, due to the fact that $P$ is a XOR-instance, her preferences can be expressed as $x_j \xor x_k=0$ or $x_j \xor x_k=1$. We have now created an instance $P'$ of \textsc{min-uncut}, and, in analogy to the proof of Theorem \ref{thm:approx-msat}, one can show that the costs of the optimal solutions of the two instances coincide. Similarly, it also holds that $\sol(P)=\sol(P')$, where 
   $\sol(P')$ corresponds to the cost of the solution of an $\alpha$-approximation algorithm for \textsc{min-uncut}, whereas $\sol(P)$ corresponds to the cost of the solution of the algorithm that transforms any XOR-instance $P$ of \cms\ to an instance $P'$ of \textsc{min-uncut}, as previously described, and then uses an $\alpha$-approximation algorithm for \textsc{min-uncut} on $P'$. Utilizing the $\log{n}$ approximation from the paper by \citeauthor{garg1996approximate} \citeyear{garg1996approximate} (Section 8 therein) for \textsc{min-uncut}, where $n$ is the number of variables in the instance, we obtain a $\log{m}$ approximation for \cms, under the discussed assumptions. 
   
   When it comes to OR-instances, it suffices to observe that a voter that casts a conditional ballot in such an instance could express his preference with a logical formula that is of the following form: $(x_j\wedge x_k)\vee(\overline{x}_j\wedge x_k)\vee (x_j\wedge \overline{x}_k)$, for some $x_j\in \{d_j,\overline{d_j}\}$ and $x_k\in \{d_k,\overline{d_k}\}$. But such an expression can be equivalently written as $x_k\vee (x_j\wedge \overline{x}_k)$ which, in turn, is equivalent to $x_k \vee x_j,$ which can be seen as a constraint that is formed by a single clause in 2-CNF form. The rest of the arguments are identical to the case of XOR-instances and the approximation factor follows from the work by \citeauthor{klein1997approximation} \citeyear{klein1997approximation} (Section 3.3 therein), where a $\log k\log\log k$ approximation algorithm is presented for the \textsc{2-cnf clause-deletion} problem, where $k$ is the number of variables in the given formula.
\end{proof}

   We note that slightly better approximation factors are possible for \cms, under both assumptions, leveraging results by \citeauthor{agarwal2005log} \citeyear{agarwal2005log}. However, this comes with the caveat of introducing randomization techniques, a debatable aspect in the context of social choice settings. Concluding this section, we highlight the attainment of a bounded approximation guarantee for every conceivable scenario, for the cases where all voters that are casting conditional ballots approve either one, two, or three combinations of values. Therefore, we have achieved positive results across the spectrum, albeit exclusively under the assumption that voters are required to approve the same number of combinations per conditional issue.

\section{Strategic Control of CMS Elections}
\label{sec:controlling}

In this section, we consider strategic aspects of \cms\ and study questions related to controlling an election of interdependent issues, which falls under the broad and well studied umbrella of influencing election outcomes. Suppose that there is an external agent (called {\it{controller}}) who has a strong preference for a specific value of some (or every) issue in a \cms\ election. 
One of the instruments for enforcing a desirable value for the issue(s) the controller cares about, is by enabling new voters to participate or by disabling some existing voters, which can be done for example by changing the criteria for eligibility of voters.
Furthermore, a controller could add more choices for the issues under consideration or delete existing ones, towards enforcing her will. We refer to the paper by \citeauthor{chen2017elections} \citeyear{chen2017elections} for related examples and further motivation. 
Finally, it is reasonable to assume that the controller does not have unlimited power, and therefore, she is capable of adding or deleting only a certain number of voters or alternatives. 

\interfootnotelinepenalty=10000

Each combination of control features (i.e., addition vs deletion, voters vs alternatives, single issue vs multiple issues) gives rise to a different control type, namely control either all or a single issue by deleting voters (\textsc{cdv}), by adding voters (\textsc{cav}), by deleting alternatives (\textsc{cda}), or by adding alternatives (\textsc{caa}). In this manner, we obtain 8 distinct algorithmic problems. Following the terminology of \citeauthor{HHR07} \citeyear{HHR07}, we say that a voting rule is \textit{vulnerable} to a certain control type, if the corresponding problem is always solvable in polynomial time. If the problem is $\mathcal{C}$-hard for a complexity class $\mathcal{C}$, we consider the rule to be \textit{resistant}\footnote{It's crucial to bear in mind that the notion of resistance is rooted in the realm of worst-case instances, supported by our NP-hardness results. These results indeed pose a barrier for controllers seeking to manipulate the outcomes, but this may not be true for every instance of the problem. In real-world scenarios, the susceptibility of \cms\ to such attempts may exhibit different behaviour.} to the specific control type (typically $\mathcal{C}$ is the class \np). In the cases where it is not possible for a controller to affect the election towards fulfilling her will, independent of complexity theory assumptions, we say that the rule is \textit{immune} to the corresponding control type. The formal definitions of the control problems appear in the following subsections and are adaptations to \cms\ elections, of the original definitions of control problems \cite{BTT92}. 
An overview of the results we obtained follows.

\begin{table*}[h!]
\resizebox{\textwidth}{!}{%
\begin{tabular}{c|ccc|ccc|ccc|}
             & \multicolumn{3}{c|}{\textbf{CDV\, \& \, CAV}} & \multicolumn{3}{c|}{\textbf{CDA}}    & \multicolumn{3}{c|}{\textbf{CAA}}    \\
             & $\Delta=0$  & $\Delta=0$ & $\Delta=1$ & $\Delta=0$ & $\Delta=1$ & $\Delta=1$ & $\Delta=0$ & $\Delta=1$ & $\Delta=2$ \\
 & $d=\bigoh{(1)}$ & $d=\omega{(1)}$ & $d=\bigoh{(1)}$ & $d=\Omega{(1)}$ & $d=\bigoh{(1)}$ & $d=\Omega{(n)}$ & $d=\Omega{(1)}$ & $d=\Omega{(n)}$ & $d=\bigoh{(1)}$ \\ \hline
\textbf{ALL} & R           & R          & R          & V          & R          & R          & I          & I          & I          \\ \hline
\textbf{1}   & V           & R          & R          & V          & V          & R          & I          & R          & R          \\ \hline
\end{tabular}}
\caption{Results on Controlling \cms\ elections. R stands for (worst-case) Resistant (i.e. \np-hard), V for Vulnerable (i.e. polynomially solvable) and I for Immune (i.e. impossible, independent of complexity theory assumptions). For a \cms\ instance on $n$ voters, we denote as $\Delta$ the maximum in-degree of every voter's dependency graph ($\Delta = \max_{i\in [n]} \Delta_i$) and $d$ the maximum domain size. 
\label{tab:control}}
\end{table*}

\subsection{Controlling Voters}
\label{subsec:voters}

We start with the problems of adding or deleting voters for enforcing a specific outcome either for a single issue or for every issue of the election. 

\vspace{0.25cm}

\noindent\fbox{%
    \parbox{0.987\columnwidth}{
\textbf{Instance:} A \cms\ election $(I,D,V\cup V',B)$, where $V$ is the set of registered voters and $V'$ is the set of yet unregistered voters with $V\cap V'=\emptyset$ (for use only by {\sc{cav}}, for {\sc{cdv}} assume that $V'=\emptyset$), an integer quota $q$, a distinguished alternative $p_j \in D_j$ for a specific issue $I_j$ or an outcome $p \in D$ (for the ``ALL" versions) specifying an alternative for every issue.\\
\textbf{Problem \cavone\ (resp. \cdvone):} Does there exist a set $V''\subseteq V'$ (resp. $V''\subseteq V$), with $|V''| \leq q$, such that $p_j$  is the value of issue $I_j$ in every optimal \cms\ solution of the profile where the only considered ballots are those from the voters of the set $V\cup V''$ (resp. $V\setminus V''$)? 
}}

\noindent\fbox{%
    \parbox{0.987\columnwidth}{
\textbf{Problem \cavall\ (resp. \cdvall):} Does there exist a set $V''\subseteq V'$ (resp. $V''\subseteq V$), with $|V''| \leq q$, such that $p$ is the unique optimal \cms\ solution of the profile where the only considered ballots are those from the voters of the set $V\cup V''$ (resp. $V\setminus V''$)?
}}

\begin{remark}
\label{rem:control-voters}
One has the option of either breaking ties in favor of the controller, if there are multiple optimal solutions in \cms\ (as done by \citeauthor{Davies+11} \citeyear{Davies+11}), or demand that the controller's will is fulfilled in every optimal outcome. We focus on the second case, as is also done in the seminal paper of \citeauthor{BTT92} \citeyear{BTT92}.
    Additionally, it is possible that the controller has a strong opinion not just for a single or all issues, but for a subset of issues. As a starting point, we have chosen to consider the two extremes (and intuitively simpler versions). 
\end{remark}

We now present our results for these 4 problems, exhibiting that it is not generally easy for a controller to enforce her will in such elections. In fact, computational hardness of controlling by adding or deleting voters can be established even for very simple forms of elections, without even the presence of conditional ballots, as shown in the two theorems that follow.

\begin{theorem}
\label{thm:cdvall}
\cdvall\ is \np-hard even for Unconditional Minisum and for a binary domain in each issue.
\end{theorem}

\begin{proof}
To prove the \np-hardness, we will have a reduction from the {\sc{vertex cover}} problem. Thus we start with an instance $(G=(V,E),k)$, which asks if there is a vertex cover of size at most $k$, and create an instance $P$ of \cdvall. 

For every edge $e \in E$, we add an issue $I_e$ having two possible alternatives, and denote its domain by $D_e = \{d_e,\overline{d_e}\}$. For every vertex $v \in V$, we add a voter voting unconditionally for $d_e$, if $e$ is incident to $v$ and being satisfied with both $\{d_e,\overline{d_e}\}$ otherwise. Let there also be $2$ dummy voters who are satisfied only with $\overline{d_e}$ for every issue $I_e$. Hence, all the ballots are unconditional, and we have an empty global dependency graph. 
For the quota parameter, we use $q=k$, and suppose that the controller wants to enforce the alternative $\overline{d_e}$ for every issue $I_e$.
This completes the description of the \cdvall\ instance, where the goal is to decide if there exists a set $V''$ of size at most $q$, such that deleting those voters enforces the controller's desirable outcome.

Suppose that there exists a vertex cover $S\subseteq V$ of $G$, of size at most $k$. Since each edge of $G$ has at least one endpoint in $S$, by removing all voters that correspond to $S$, each alternative $d_e$ loses at least one approval vote. Hence, $d_e$ would cause two dissatisfactions to the dummy voters (the others are indifferent), whereas $\overline{d_e}$ causes at most one dissatisfaction. Therefore, selecting the alternative $\overline{d_e}$ for every issue $I_e$ is the unique optimal solution,

For the reverse direction, suppose there exists a set of voters $S$, whose removal causes the outcome $(\overline{d_e})_{e\in E}$ to become the unique optimal solution. First, we may assume that $S$ does not contain any of the dummy voters (otherwise, add them back to the instance, and the total dissatisfaction score will not be affected). Suppose that $S$ is not a vertex cover in $G$, and that at least one edge $e$ is not covered by $S$. But this means that the removal of $S$ from the \cdvall\ instance will leave intact the two voters that are satisfied only with $d_e$, and therefore $d_e$ can also be selected in an optimal solution (it causes the same number of dissatisfactions as $\overline{d_e}$). This contradicts the fact that the removal of $S$ resulted in the unique optimal solution with $\overline{d_e}$ selected for every issue $I_e$. 
\end{proof}

\begin{theorem}
\label{thm:cavall}
\cavall\ is \np-hard even for Unconditional Minisum and for a binary domain in each issue.
\end{theorem}
\begin{proof}
    The proof is a simple adaptation of a reduction given for almost the same problem but in the context of the classic (unconditional) approval voting rule \cite{HHR07}. For the sake of completeness, we provide the full construction here. We stress that we cannot directly establish \np-hardness by applying the result of that work because when there are no conditional ballots, the version of approval voting as defined there selects as winner(s) the candidates who have the highest number of approvals, whereas Unconditional Minisum selects only candidates who are approved by at least 50\% of the voters. In the instances used in the reductions by \citeauthor{HHR07} \citeyear{HHR07} (see Theorem 4.43 therein), there are losing candidates who are approved by more than 50\% of the voters, hence their proofs do not apply directly.  

We start with an instance $P$ of {\sc{exact-3-cover (x3c)}} where $B = \{b_1,\dots, b_m\}$ with $m=3k$ is the universe, and ${\cal F} = \{S_1,\dots, S_n\}$ is a collection of sets with $|S_i|=3$, for every set $S_i$. The goal is to decide if there is an exact cover, i.e. a subcollection of sets from ${\cal F}$ such that each element of the universe belongs to exactly one of these sets.

We now define a \cms\ election where the set of issues is $I = B\cup \{I_{m+1}\}$ and each issue has a binary domain, with $D_j = \{b_j, \overline{b_j}\}$ for $j\in [m]$, and $D_{m+1} = \{w, \overline{w}\}$. The set of voters is as follows:
\begin{itemize}
    \item There are $k-2$ registered voters who are satisfied with $b_j$ for $j\in [m]$, and with $\overline{w}$. They are dissatisfied with the complements of these alternatives. 
    \item There is one registered voter who is satisfied only with $\overline{b_j}$ for $j\in [m]$ and with $\overline{w}$.
    \item There are $n$ unregistered voters corresponding to the sets of {\sc{x3c}} instance. The voter corresponding to $S_i$ is satisfied only with the 3 alternatives of $S_i$, and with $w$.
\end{itemize}
To finish the description, we set the quota parameter $q$ equal to $k$ and the desirable outcome of the controller to be $(\overline{b_1},\dots, \overline{b_m}, w)$. Hence, the goal in the \cavall\ instance is to decide if there exists a set of unregistered voters $V''$ with $|V''|\leq k$ such that adding $V''$ to the registered voters makes the desirable outcome the unique optimal solution.

Suppose now that there exists an exact cover in $P$. Since $m=3k$, the cover consists of exactly $k$ sets. Select as $V''$ the $k$ unregistered voters corresponding to the cover. We now have a total of $2k-1$ voters in the election. For the first $m$ issues, the alternative $b_j$ satisfies exactly $k-1$ voters and dissatisfies $k$ voters, hence the optimal solution selects $\overline{b_j}$ for $j\in [m]$. For the last issue, the value $w$ satisfies $k$ voters and dissatisfies the remaining $k-1$ voters. Hence, the unique optimal solution when adding the set $V''$ is precisely $(\overline{b_1},\dots, \overline{b_m}, w)$.

For the opposite direction, suppose that there is a set $V''$ of unregistered voters, with $|V''|\leq k$, such that when adding them to the registered voters, the unique optimal solution is the controller's desirable outcome. First notice that this implies that $|V''| = k$, otherwise there is not enough support for $w$ to be selected. The only other possibility would be to have $|V''| = k-1$, but then we have a tie, and there would be more optimal solutions with $\overline{w}$ instead of $w$. Since for the other issues, each $b_j$ already has a support by $k-2$ registered voters, then none of them received a support by two or more of the added voters. But these voters express a support for a total of $3k=m$ such alternatives, therefore, each $b_j$ for $j\in [m]$, receives support by exactly one of the added voters. 
\end{proof}

The next step is to see whether the hardness results of Theorems \ref{thm:cdvall} and \ref{thm:cavall} go through when the controller wishes to control just a single issue. 
For Unconditional Minisum this is not the case if we insist on a constant domain size for the designated issue.
The reason is that this can be reduced to an FPT version of the well known Set MultiCover problem. This has been already proven by \citeauthor{Bredereck+20} \citeyear{Bredereck+20} and results to the following:

\begin{proposition}[\citeauthor{Bredereck+20} \citeyear{Bredereck+20}]
\label{obs:cav-cdv}
\cavone\ and \cdvone\ can be solved in polynomial time for Unconditional Minisum if the domain size of each issue is constant. 
\end{proposition}

As a consequence, any potential hardness result for \cavone\ and \cdvone\ would have to consider either non-constant domain sizes or conditional ballots. Indeed, we establish that either of these settings suffices to establish \np-hardness. We start with elections where at least one issue has a non-constant domain size.

\begin{theorem}
\label{thm:cav-cdv-delta=0}
$\cavone$ and $\cdvone$ are \np-hard, even for Unconditional Minisum, and even if all but one issue have constant domain size.
\end{theorem}

\begin{proof}

We will only describe the proof of \np-hardness for \cavone\ and the same can be established for \cdvone\ in a very similar fashion, using almost the same reduction.

We will have a reduction from the problem of controlling a classic approval voting election by adding voters, known to be \np-hard \cite{HHR07}. We recall that in an approval voting election, voters express their approved set of candidates, and the winner (or winners in case of ties) is the candidate with the highest number of approvals. The control problem there is to ensure that a designated candidate is the unique winner of the election. Our reduction starts with an instance $P$ of the control problem in approval voting, where $V$ and $V'$ are the registered and unregistered sets of voters respectively, $p$ is a designated candidate, and $q$ is a quota. The goal is to select a set $V'' \subseteq V'$ with $|V''| \leq q$, so that the approval voting rule, when run on the voters in $V\cup V''$ will select $p$ as the unique winner. 

We create an instance $P'$ of \cavone\ where the sets of voters, registered and unregistered, are the same as in $P$. If the number of candidates in $P$ is $m$, we create a single issue in $P'$ whose domain has exactly $m$ possible alternatives, and $p$ is the designated alternative that the controller wants to promote in $P'$. For every voter in $P$ (whether coming from $V$ or $V'$), the corresponding voter in $P'$ specifies an unconditional ballot on the single issue, containing only her approved options in $P$. We also use the same quota parameter $q$ as in $P$. This completes the description of $P'$, which can be clearly constructed in polynomial time.

It is now easy to see that there exists a set $V''\subseteq V'$ of at most $q$ voters so as to ensure that $p$ will be the outcome on the single issue of $P'$, using the \cms\ rule for the voters of $V\cup V''$, if and only if the same set of voters can ensure that $p$ will be the unique winner in the approval voting election of $P$. Indeed, if the \cms\ rule, run on the voters of $V\cup V''$, selects the outcome $p$ in the instance $P'$, this means by the definition of the \cms\ rule that $p$ causes the minimum number of dissatisfactions among all possible alternatives, i.e., it has the highest number of approvals. This directly yields that $p$ will be the unique winner in the instance $P$. The reverse direction is easy to see as well, with the same reasoning. 
\end{proof}

We now study the hardness of these problems when we have conditional ballots. As the next theorem shows, it suffices to consider only profiles where each issue may depend on at most one other issue.

\begin{theorem}
\label{thm:cav-cdv-binary}
\cdvone\ and \cavone\ are \np-hard, when $\Delta\leq 1$, even for a binary domain in every issue.
\end{theorem}

\begin{proof}

We will focus on proving the statement for \cdvone. The proof for \cavone\ is very similar, requiring minimal changes, that we discuss at the end. We will prove that \cdvone\ with a binary domain for every issue, and with $\Delta=1$ is \np-hard, using a reduction from the \np-hard version of \cdvone\ with $\Delta=0$ and non-constant domains (Theorem \ref{thm:cav-cdv-delta=0}). 
We first remind the reader that the proof of Theorem \ref{thm:cav-cdv-delta=0} indicates that the problem \cdvone\ with $\Delta=0$ and non-constant domain sizes is \np-hard, even for the family of instances with just a single issue (that has a non-constant number $m$ of different alternatives), where every voter casts only approval ballots for a subset of alternatives.

Our reduction starts from an instance $P$ of \cdvone\ on a single issue with non-constant domain size (obviously $\Delta=0$, since we have only one issue), and creates an instance $P'$ of \cdvone\ with binary domain for every issue and with $\Delta=1$. 
Say that the issue of $P$ has the following $m$ different alternatives: $\{d_1,d_2,\dots, d_m\}$. Then the instance $P'$ consists of $m$ different binary issues $I_1,I_2,\dots,I_m$ such that the alternatives of issue $I_j$ are $\{d_j,\overline{d_j}\}$, for $j\in [m]$. Hence, the idea is that each alternative of the single issue of $P$ now corresponds to a different issue in $P'$ with positive and negative alternatives.  
Furthermore, if $q$ was the quota in $P$, we will use the same quota in $P'$. Finally, if $d_\ell$ was the designated alternative in $P$, for some specific $\ell\leq m$, then we will have that $d_\ell$ is the designated alternative for issue $I_\ell$ in $P'$. 

We describe now the set of voters in $P'$ as well as their preferences. 
For every voter $v$ of $P$, we add a voter $v'$ in $P'$, such that for any $j\in[m]$, if $v$ was approving the alternative $d_j$  for the single issue of $P$, then $v'$ approves $d_j$ concerning the issue $I_j$ in $P'$, otherwise, if $v$ was not approving $d_j$ in $P$, then $v'$ is indifferent in $P'$ and votes for $\{d_j,\overline{d_j}\}$. 
We also add a set of $m(m-1)L$ dummy voters, where $L=nm+q+1$ and $n$ is the number of voters in $P$. 
In particular, for every ordered pair of distinct alternatives of $P$, i.e., for every $(k,j)$, with $k, j\in [m]$, and $k\neq j$, we include $L$ voters voting $\overline{d_k}$ for issue $I_k$, $\{d_k:\overline{d_j}\}$ for issue $I_j$, and $\{d_t,\overline{d_t}\}$, for every other issue $I_t$, with $t \in [m]\setminus\{k,j\}$.

We will first prove the following claim, which is enforced by the construction of $P'$.
\begin{claim}
\label{cl:at-most-one}
Let $P''$ be the conditional approval voting profile that is derived by the deletion of a set of at most $q$ non-dummy voters from the instance $P'$. Then, in any optimal \cms\ solution of $P''$, there is exactly one issue $I_j$ for which the selected alternative is $d_j$, and for every $k\neq j$, the selected alternative will be $\overline{d_k}$.
\end{claim}
\begin{proofof}{Proof of Claim \ref{cl:at-most-one}}
Let us define first the set of outcomes where exactly one issue takes a positive value, i.e., let $POS_1=\{\{y_1,y_2,\dots,y_m\}: \exists i\in [m]: y_i=d_i, \forall j \in [m]\setminus\{i\}: y_j=\overline{d_j}\}$. We will first prove that for any solution that belongs to $POS_1$, the total dissatisfaction incurred by the set of dummy voters equals $m(m-1)L$. To prove that, we inspect an arbitrary outcome of $POS_1$, say $\{\overline{d_1},\overline{d_2},\dots,\overline{d_{p-1}},d_p,\overline{d_{p+1}},\dots,\overline{d_m}\}$, for some $p\in [m]$. It is convenient to view the set of $m(m-1)L$ dummy voters of $P''$, as being partitioned in the following 3 sets:
\begin{itemize}
    \item The $(m-1)(m-2)L$ dummy voters whose dependency graph consists of an edge $(I_k,I_j)$, such that $k,j\neq p$. These voters are dissatisfied only with respect to issue $I_j$ since they are voting for $\{d_k:\overline{d_j}\}$ and $d_k$ is not selected. They are satisfied with respect to $I_k$ since they are voting in favor of $\overline{d_k}$, which is elected, and they are indifferent (hence satisfied) with respect to all other issues.
    \item The $(m-1)L$ dummy voters whose dependency graph consists of an edge $(I_p,I_j)$, such that $j\neq p$. These voters are dissatisfied only with respect to issue $I_p$ since they are voting for $\overline{d_p}$ but $d_p$ is elected. They are satisfied with respect to $I_j$ since they are voting in favor of $\{d_p:\overline{d_j}\}$ and both $d_p$ and $\overline{d_j}$ are elected. Finally they are satisfied with respect to any $I_t$, for $t\neq p,j$, since they are indifferent for these issues.
    \item The $(m-1)L$ dummy voters whose dependency graph consists of an edge $(I_j,I_p)$ such that $j\neq p$. These voters are dissatisfied only with respect to issue $I_p$ since they are voting for $\{d_j:\overline{d_p}\}$. Furthermore they are satisfied with respect to $I_j$ since they are voting in favor of $\overline{d_j}$, and they are satisfied with respect to any other issue. 
\end{itemize}
Hence, the total dissatisfaction score of any outcome in $POS_1$ is $m(m-1)L$, due to the dummy voters. To count the dissatisfaction from the remaining non-dummy voters, we define the following quantity: let $x_i$ be the number of voters in the original instance $P$ (after deleting the voters that correspond to the ones deleted in $P'$) who had $d_i$ in their approval list in $P$. Then, it can be verified that the total dissatisfaction due to the non-dummy voters for the profile  $P''$, is $\sum_{i\in[m]\setminus p}x_i$. 
Hence, to concude, the total dissatisfaction score for $P''$, under any outcome that belongs to $POS_1$ is equal to $m(m-1)L+\sum_{i\in[m]\setminus p}x_i$.

We will now compare the dissatisfaction score of the outcomes in $POS_1$ with all other possible outcomes. We can define analogously the set of outcomes $POS_{2}=\{(y_1,y_2,\dots,y_m): \exists i,j\in [m]: y_i=d_i,y_j=d_j, \forall k \in [m]\setminus\{i,j\}: y_k=\overline{d_k}\}$. We will first prove that the total dissatisfaction incurred by dummy voters in outcomes from $POS_2$ is at least $m(m-1)L+2L$. To prove it, we inspect an arbitrary outcome of $POS_2$, say $(y_1, \dots, y_m)$ with $y_p = d_p$, $y_r = d_r$, for some specific $p, r$, and also $y_j = \overline{d_j}$, for any $j\neq p, r$. 
We analyze the set of dummy voters, by partitioning them in 4 sets as follows: 
\begin{itemize}
    \item The $(m-2)(m-3)L$ dummy voters whose dependency graph has the edge $(I_k,I_j)$, for some $k,j$ where both $k, j\neq p,r$. These voters are dissatisfied only with respect to issue $I_j$ since they are voting for $\{d_k:\overline{d_j}\}$ but $\overline{d_k}$ is elected. They are satisfied with respect to $I_k$ since they are voting in favor of $\overline{d_k}$. Finally they are indifferent and hence satisfied with respect to all other issues.
    \item The $2(m-2)L$ dummy voters whose dependency graph has either the edge $(I_p,I_j)$ or $(I_r,I_j))$, with $j\neq p,r$. We analyze the ones with the edge $(I_p,I_j)$, and the same conclusion holds for the other case as well. These voters are dissatisfied only with respect to issue $I_p$ since they are voting for $\overline{d_p}$, but $d_p$ is elected. They are satisfied with respect to $I_j$ since they are voting in favor of $\{d_p:\overline{d_j}\}$ and both $d_p$ and $\overline{d_j}$ are elected. Finally they are satisfied with respect to any $I_t$, for $t\neq p,j$, since they are indifferent.
    \item The $2(m-2)L$ dummy voters whose dependency graph consists of the edge $(I_j,I_p)$ or $(I_i,I_r)$ for some $j\neq p,r$. We argue about the voters with the edge $(I_j,I_p)$ as the other case is also identical. These voters are dissatisfied only with respect to issue $I_p$, since they are voting for $\{d_j:\overline{d_p}\}$ but $d_p$ is elected. They are satisfied with respect to $I_j$ since they are voting in favor of $\overline{d_j}$ which is elected. Finally they are indifferent with respect to other issues.
    \item The $2L$ dummy voters whose dependency graph consists of the edge $(I_p,I_r)$ or $(I_r,I_p)$. These voters are dissatisfied with respect to issues $I_r$ and $I_p$. Consider the ones with the edge $(I_p,I_r)$. They are voting for $\{d_p:\overline{d_r}\}$ but $d_r$ is elected. Additionally, they are voting for $\overline{d_p}$ but $d_p$ is elected. They are satisfied with respect to any other issue since they are indifferent.
\end{itemize}

The above analysis shows that the dissatisfaction score due to the dummy voters is $(m-2)(m-3)L+2(m-2)L+2(m-2)L+4L=m(m-1)L+2L$. This is larger by the term $2L$, compared to the dissatisfaction of the dummy voters in outcomes of $POS_1$. By the choice of $L$, it is impossible that the dissatisfaction of the non-dummy voters causes an outcome of $POS_2$ to have a better or equal score than those of $POS_1$ (note that $L> mn$ and the total dissatisfaction of the non-dummy voters is bounded by $mn$). Hence, the optimal solution cannot be attained by $POS_2$. 
In a similar manner, we can define $POS_{k}$, for any $k>2$, and prove that the total dissatisfaction score of any outcome in $POS_1$ is less than the dissatisfaction score of any outcome in $POS_{k}$. Equivalently, there can be no more than a single positive alternative in the optimal outcome of $P''$. 

Finally, we also need to compare with the ``all-negatives" outcome, where $\overline{d_j}$ is selected for every $j\in [m]$.
It is a matter of a simple case analysis (like the ones before) to prove that the dissatisfaction incurred by the dummy voters equals $m(m-1)L$, which equals the dissatisfaction of the dummy voters for outcomes in $POS_1$ as well. Hence, now we only need to argue about the non-dummy voters; it is safe to say that these are more than $q$. To do this, note that in the original instance $P$, we can assume without loss of generality that at least one voter approves at least one alternative, otherwise we have a trivial election, where no one expresses any preferences. This means that $x_p>0$ for at least one issue $p\in [m]$. By the construction of the preferences for the non-dummy voters in $P'$, the dissatisfaction score for the ``all-negatives" outcome equals $\sum_{i \in [m]}x_i$. Recall also that for the outcome in $POS_1$, where only issue $p$ has a positive value, the dissatisfaction of the non-dummy voters is $\sum_{i \in [m]\setminus\{p\}}x_i$. Therefore, we have that there exists an outcome in $POS_1$ which causes strictly less dissatisfaction to the electorate, and this concludes the proof.
\end{proofof}

Let $P$ be a yes instance, i.e., say that there is a set $S$ of at most $q$ voters, the deletion of which causes the election of $d_{\ell}$ in $P$, as the unique winner. Note that for every voter in $P$ there is a corresponding voter in $P'$. Consider the deletion of the set $S'$ from $P'$ that corresponds exactly to the voters of $S$ from $P$. By exploiting Claim \ref{cl:at-most-one}, it suffices to prove that by deleting $S'$, the outcome $p_\ell=\{ \overline{d_1},\overline{d_2},\dots,\overline{d_{\ell-1}},d_\ell, \overline{d_{\ell+1}},\dots, \overline{d_{m}}\}$ causes strictly less dissatisfactions in the electorate than $p_j=\{ \overline{d_1},\overline{d_2},\dots,\overline{d_{j-1}},d_j, \overline{d_{j+1}},\dots, \overline{d_{m}}\}$, for any $j\neq \ell$. To do so, for any $i\in [m]$, let $x_i$ be the number of voters (among the remaining ones, after deleting the set $S$) who had $d_i$ in their approval list in $P$. By using the same argument as in the proof of Claim \ref{cl:at-most-one}, we can see that for the non-dummy voters of $P'$ who correspond to those of $P$, the number of dissatisfactions caused by the outcome $p_\ell$ in $P'$ is $(\sum_{i \in [m]\setminus \ell,j}x_i)+x_j$. Also, by doing the same counting argument as in the first part of the proof of Claim \ref{cl:at-most-one}, each dummy voter will be dissatisfied with exactly one issue, and hence they contribute a total of $m(m-1)L$ in the cumulative number of dissatisfactions. 
In the same manner, we can also have that the total number of dissatisfactions caused by the outcome $p_j$ is $(\sum_{i \in [m]\setminus \{l,j\}}x_i)+x_\ell + m(m-1)L$. 
Given that $d_\ell$ was the winning alternative in $P$ after the deletion of voters, it is true that $x_\ell>x_j$ and hence the dissatisfaction caused by $p_j$ is greater than the dissatisfaction caused by $d_\ell$. By Claim \ref{cl:at-most-one}, this concludes the first direction of the proof.

For the reverse direction, say that there is a set of voters $S'$ in $P'$, after the deletion of which, $d_\ell$ is the winning alternative for issue $I_{\ell}$. 
We note that because of the large value of $L$, including any dummy voter in the set $S'$ will not influence the final outcome (since $L>q,$ even if all the deleted voters are dummy ones, we cannot enforce a different outcome than the outcome without deletions). Hence, we can assume that the set $S'$ has no dummy voters.
We can choose then to delete the corresponding set of voters $S$ in $P$ and we will need to prove that, in that case, the elected alternative will be $d_\ell$ for the single issue of $P$. By Claim \ref{cl:at-most-one}, and since we assumed that $d_\ell$ is selected for issue $I_\ell$, we know that the optimal solution in the instance after the deletion of $S'$ selected $\overline{d_j}$ for any other $j\neq \ell$. Thus, the winning outcome is $p_\ell=\{ \overline{d_1},\overline{d_2},\dots,\overline{d_{\ell-1}},d_\ell, \overline{d_{\ell+1}},\dots, \overline{d_{m}}\}$.
Therefore, the number of dissatisfactions caused by $p_\ell$ is lower than the number of dissatisfactions caused by $p_j=\{ \overline{d_1},\overline{d_2},\dots,\overline{d_{j-1}},d_j, \overline{d_{j+1}},\dots, \overline{d_{m}}\}$, for any $j\neq \ell$. But this implies that $(\sum_{i \in [m]\setminus\{\ell,j\}}x_i)+x_j<(\sum_{i \in [m]\setminus\{\ell,j\}}x_i)+x_{\ell}$, or equivalently, $x_\ell>x_j$, which means that the number of voters that are not included in $S$, and have $d_\ell$ in their approval list, is greater than the number of voters not included in $S$, who have $d_j$ in their list for any $j\neq \ell$. Thus, in $P$, $d_\ell$ will be selected, fulfilling the controller's will.

\noindent {\it Adjustments for the \cavone\ reduction:} So far we have established that \cdvone\ is \np-hard. To prove that \cavone\ is also \np-hard for binary domains, we will perform a similar reduction, this time from \cavone\ with a single issue and with a non-constant domain size (which is again \np-hard by Theorem \ref{thm:cav-cdv-delta=0}). The construction is almost the same, with the difference being that in \cavone, there are both registered and unregistered voters. Our reduction will assign preferences in the same way as in the proof for \cdvone\ and will simply maintain the separation into registered and unregistered voters in the created instance $P'$. Furthermore, we include the dummy voters in the set of registered voters and we set $L=m(n+q)+1$.
Everything else in the reduction is the same as before, and it is a matter of calculations analogous to the proof of \cdvone\ to verify the correctness of the reduction.  
\end{proof}

To conclude this subsection, we have now a complete picture for the level of resilience against the malicious actions of adding or deleting voters. Our results 
act in favor of the \cms\ rule, showing that a potential controller cannot easily (in terms of computational complexity) enforce her own desirable outcomes, apart from a single case, as shown in Table \ref{tab:control}. Interestingly, this, single, polynomially solvable case concerns the unconditional setting and when one moves to the conditional case, the considered problem becomes hard for the controller. Finally, it's worth noting that despite the need for distinct proof approaches in certain cases, the results for the two examined problems (deleting/adding voters) are identical; this observation does not extend to the problems related to the deletion and addition of alternatives, as elucidated in the forthcoming subsection.

\subsection{Controlling Alternatives}
\label{subsec:alt}

We now consider the analogous control problems, regarding the addition or deletion of alternatives, instead of voters. It turns out that the picture, from the computational complexity viewpoint, differs sufficiently from the problems considered in the previous subsection.

\vspace{0.25cm}

\noindent\fbox{%
    \parbox{0.968\columnwidth}{
\textbf{Instance:} A \cms\ election $(I,D,V,B)$, where $D =(D_1\cup D_1')\times \dots \times (D_m\cup D_m')$ such that $D_k$ is the set of qualified alternatives of each issue $I_k$ and $D_k'$ (where $D_k\cap D_k'=\emptyset$) is a set of spoiler alternatives for each $I_k$ (for use only by {\sc{caa}}, for {\sc{cda}} we assume $D_k'=\emptyset$), an integer quota $q$, a distinguished alternative $p_j \in D_j$ for a specific issue $I_j$ or an outcome $p \in \bigtimes_{k\in [m]} D_k,$ specifying an alternative for every issue.\\
\textbf{Problem \caaone\ (resp. \cdaone):} Does there exist a set $D'' \subseteq \cup_{k\in[m]} D_k'$ (resp. $D'' \subset \cup_{k\in[m]} D_k$), with $|D''| \leq q$, such that $p_j$ is the value of the issue $I_j$ in every optimal \cms\ solution of the profile where the domain $D_k$ of each issue $I_k$ is enlarged by the alternatives in $D''\cap D_k'$ (resp. reduced by the alternatives in $D''\cap D_k$)? 
\\
\textbf{Problem \cdaall:} Does there exist a set $D'' \subset \cup_{k\in[m]} D_k$, with $|D''| \leq q$, such that $p$ is the unique optimal \cms\ solution of the profile where the domain $D_k$ of each issue $I_k$ is reduced by the alternatives in $D''\cap D_k$?\\
\textbf{Note:} For \cdaone\ and \cdaall, we also require that for every $k$, $|D_k \setminus D''|\geq 1$.
}
}

\begin{remark}
We first note that all the comments made in Remark \ref{rem:control-voters} are applicable here as well.
Also, we have not included \caaall\ in our definitions as \cms\ is trivially immune to adding spoiler alternatives in order to enforce a qualified alternative in every issue.
Concerning the problem \caaone, we assume that the voters in $B$ may express an opinion about any outcome of every issue, whether it is a qualified one or a spoiler. Additionally, another way to define such problems would be to allow the controller to completely delete or add issues instead of just alternatives. However, given the existence of dependency graphs, erasing an issue can make the preferences of a voter ill-defined. 
    Lastly, the constraint that $|D_k \setminus D''|\geq 1$, for \cdaone\ and \cdaall, is to ensure that the controller cannot eliminate all the alternatives of an issue.
\end{remark}

\begin{proposition}
Unconditional Minisum, with arbitrary domain size is immune to \caaone. For the same setting, \cdaone\ and \cdaall\ can be solved in polynomial time. 
\end{proposition}

\begin{proof}
To solve \cdaone\ and \cdaall\ we only have to observe that to control a single issue by deleting alternatives in the unconditional case, one can check if the quota is large enough to delete all alternatives that achieve higher approval score than the designated one(s). At what concerns \caaone, the definition of immunity directly applies, since the controller cannot enforce a designated alternative in Unconditional Minisum by adding some other alternative (whether for the same or for a different issue).
\end{proof}

When we move to instances with conditional ballots, the problems \cdaone\ and \cdaall\ do become hard (with the exception of Proposition \ref{prop:cdaone}). We start with the hardness of \cdaall.

\begin{theorem}
\label{cdaall-deg1}
\cdaall\ is \np-hard, when $\Delta\leq 1$, and even for a binary domain size in every issue.
\end{theorem}

\begin{proof}
Let $P= (G,k)$ be an instance of \textsc{vertex cover}, for an undirected graph $G$ with $n$ vertices and $m$ edges, and an upper bound $k$ on the desirable size of the cover. We denote by $e_1,e_2,\dots, e_m$, the edges of the graph. We will present a reduction from $P$ to an instance $P'$ of \cdaall. Let there be the following $2m(k+1)+n$ issues:
\begin{itemize}
    \item For every edge $e_\ell$ of $G$, $\ell\in [m]$, and for $j\in [k+1]$ let $I_{\ell,j}$ and $I'_{\ell,j}$ be a pair of binary issues of domain $\{e_{\ell,j},\overline{e_{\ell,j}}\}$ and $\{e'_{\ell,j},\overline{e'_{\ell,j}}\}$ respectively,. We refer to these issues as edge issues.
    \item For every vertex $v$ of $G$ let $I_v$ be a binary issue of domain $\{d_v,\overline{d_v}\}.$ We refer to these issues as vertex issues.
\end{itemize}

Note that, we have added one vertex issue for every vertex of $G$ and, essentially, we have introduced two edge issues for every edge of $G$, but in $(k+1)$ 'copies'. This will play a significant role in the reverse direction of the reduction. The voters of the created instance $P'$ will be $2m+3$ in total, and we partition them as follows:
\begin{itemize}
    \item Group 1: There are $2$ edge voters for every edge $e_i$ of $G$. For an edge $e_i=(u,v)$ and for every $j\in[k+1]$, these voters submit the same ballot $\{d_u:e_{i,j},d_v:e'_{i,j}\}$ (where we have arbitrarily chosen that the satisfaction of one edge issue depends on the vertex issue $I_u$ and the other edge issue depends on the remaining vertex issue $I_v$). The voters are indifferent for the rest of the issues. Hence, every such voter is interested in exactly $2(k+1)$ edge issues, each conditioned on either $I_u$ or $I_v$. At what follows, we will refer to these voters as ``group 1" or edge voters.
    \item Group 2: There are 2 voters who are voting for $\{\overline{\strut e_{i,j}}:\overline{\strut e'_{i,j}}\}$, for every pair $(i,j)\in [m]\times[k+1]$, and who are indifferent for every other issue. At what follows, we will refer to these voters as ``group 2".
    \item There is one voter, that we will refer to as the special voter, who is voting unconditionally $\overline{d_1},\overline{d_2},\dots,\overline{d_n}$ for the vertex issues, and is indifferent for every other issue.
\end{itemize} 
 To complete the construction of the instance $P'$, we use $k$ as the quota parameter, and we suppose the controller wants to enforce the outcome $\{ (\overline{e_{i,j}},\overline{e'_{i,j}})_{i\in[m], j\in[k+1]}, (\overline{d_j})_{j\in[n]}\}$ 
 by removing at most $k$ alternatives. It is trivial to observe that all the issues in the instance $P'$ are of binary domain and that for every voter $i$, $\Delta_i\leq 1$ in her dependency graph. 
 
Before we proceed, we note that the designated outcome fully satisfies the voters from group 2 as well as the special voter, however, it dissatisfies every edge voter with respect to all $2(k+1)$ issues they care about each, hence it produces a total dissatisfaction score of $4m(k+1)$. Additionally, we highlight that there are also other outcomes with the same score, which prevent the designated one from being the unique winner without deleting any alternatives in $P'$. 
For instance, if $e_t$ is the edge $(u,v)$ of $G$, then consider the following outcome, for any $\ell\in [k+1]$:
 \begin{equation}
     \label{problematic}
     \{ (e_{t,\ell},e'_{t,\ell}), (\overline{e_{i,j}},\overline{e'_{i,j}})_{(i,j)\in[m]\times[k+1]\setminus (t,\ell)}, du, dv, (\overline{d_j})_{j\in[n]\setminus\{u,v\}}\}
 \end{equation}

The outcome described in Equation \ref{problematic} has a dissatisfaction score of $4(m-1)(k+1)+4k$ from group 1 and a dissatisfaction score of $1$ from each voter in group 2 and a dissatisfaction score of $2$ for each remaining voter, leading to a total score of $4m(k+1)$.

To see now the forward direction of the reduction, suppose there exists a vertex cover in $G$ of size at most $k$, which is formed say by a set of vertices $S\subset [n]$. Then we choose to delete in the created instance $P'$ of \cdaall, the corresponding positive alternatives $\{d_j\}$ for $j\in S$, from the vertex issues $\{I_j\}_{j\in S}$, respectively. Hence, any solution to the resulting instance after these deletions, should now definitely contain the alternative $\overline{d_j}$ for every issue $I_j,$ such that $j\in S$. Clearly, the designated outcome still remains a valid solution with a total score of $4m(k+1)$. We need to see what happens with the rest of the possible outcomes. 

One can easily verify that the best solution among the feasible ones in which all edge voters are dissatisfied with respect to all the issues they care about, is precisely the designated outcome (because it satisfies in all issues all the other voters from group 2 as well as the special voter, and also no other solution can achieve the same). 
Hence, it remains to see if there exists any optimal solution with at least one pair of edge voters  satisfied with respect to at least one issue. 

For the sake of contradiction, assume that this is the case, and consider a pair of such edge voters, corresponding to edge $e = (u,v)$. Without loss of generality, suppose that $u$ does not belong to the vertex cover, and hence $d_u$ is an available alternative, that has been selected in the optimal solution we are considering, and also that $d_v$ has been deleted. This is due to the vertex cover property, implying that either $d_u$ or $d_v$ has been deleted and furthermore, if both had been deleted, the pair of edge voters that we focus on, would not be satisfied with respect to any issue, contrary to what we have assumed. The selection of $d_u$ may also cause other edge voters (whose edge is incident with $u$) to be satisfied as well, with respect to some issues. To arrive at a contradiction, and for ease of notation, let $e_1,\dots, e_\ell$, be the edges that have $u$ as an endpoint, and that correspond to edge voters who are satisfied with respect to at least one of the issues they care about, under the optimal solution we are considering. Additionally, for such an edge $e_i$, let $m(i)$ be the number of issues that correspond to $e_i$, and with respect to which, the two edge voters of $e_i$ are satisfied (this occurs because the edge voters may have declared $d_u:e_{i, j}$ or $d_u:e'_{i, j}$). Clearly $m(i)\leq k+1$. Counting the total number of satisfactions of edge voters due to the selection of $d_u$, we get exactly $2\sum_{i=1}^{\ell}m(i)$. 
But if we now replace $d_u$ by $\overline{d_u}$ and set all the edge issues of the edges $e_1,\dots, e_\ell$, to their negative value, then we will dissatisfy all the edge voters that we were satisfying before, but we get $2\sum_{i=1}^{\ell}m(i)$ new satisfactions from group 2 and one new satisfaction from the special voter. 
Therefore, we reach an outcome with a lower  dissatisfaction score, contradicting the fact that we started with an optimal solution. 

We conclude that there cannot exist an optimal solution, after the deletions we made, where some of the voters from group 1 enjoy any satisfaction. Therefore, we can only satisfy group 2 and the special voter, and we can conclude that after the deletion of at most $k$ (positive) alternatives that correspond to a vertex cover, the designated outcome becomes the unique winner.

For the reverse direction, suppose that there is a set $D$ of at most $k$ alternatives, the deletion of which, forces the designated outcome to be the unique optimal solution. Trivially, $D$ cannot contain negative values neither from edge nor from vertex issues. We denote by $D_V$ the subset of $D$ that contains positive values from vertex issues and let $S$ be the corresponding set of vertices in $G$. We claim that $S$ forms a vertex cover of $G$. Towards a contradiction, say that $S$ is not a vertex cover. Then, there exists an edge $e_t=(u,v)$ such that both $d_u,d_v\notin D_V$, thus $d_u$ and $d_v$ are still feasible alternatives, after the deletions we have made. 
Note that due to the budget constraint, it holds that $|D''|\leq k$. Hence out of the $2(k+1)$ issues that the edge voters corresponding to $e_t$ care about, there exists an index $\ell$, such that both $e_{t,\ell}$ and $e'_{t,\ell}$ are still available, after the deletion of $D$. But then, the outcome described in Equation \eqref{problematic} is still feasible, which contradicts the fact that the designated outcome became the unique winner after the deletion of $D$. 
\end{proof}

Moving to \cdaone\ and \caaone\, we show that we can have hardness results, but only under a non-constant domain size for at least one issue. The proof of Theorem \ref{thm:cdaone-largedomain} below, shows a connection with some natural problems on graphs, that have been previously linked to election control for other voting rules \cite{BU09}.

\begin{theorem}
\label{thm:cdaone-largedomain}
\caaone\ and \cdaone\ are \np-hard, when $\Delta\leq 1$, and even when the treewidth of the global dependency graph is at most one, but with non-constant domain size in at least one issue.
\end{theorem}
\begin{proof}
We will first prove the hardness of \cdaone. We will perform a reduction from the \np-hard problem {\textsc{max out-degree deletion (mod)}} \cite{BU09}. 

\vspace{0.25cm}

\noindent\fbox{%
    \parbox{0.974\columnwidth}{
\textbf{Instance:} A directed graph $G=(V,E)$, a special vertex $p\in V$ and an integer $k\geq 1$.\\
\textbf{Output:} Does there exist $V'\subseteq V$ with $|V'|\leq k$ such that $p$ is the only vertex of maximum out-degree in $G[V\setminus V']$? 
}}

\vspace{0.25cm}

For $S\subseteq V$, we denote by $deg_S(u)$ the out-degree of vertex $u$ in a graph $G=(V, E)$, when we count only outgoing edges towards the vertices of $S$. Let $P= (G,p,k)$ be an instance of \textsc{mod} in a directed graph with $n$ vertices and $m$ edges
We create a \cdaone\ instance, where we have one issue $I_j$ for every vertex $v_j, j\in [n]$ and an extra issue $I_0$, hence $I= \{I_0, I_1,I_2,\dots,I_n\}$. For $j\in [n]$, the domain of issue $I_j$ is binary in the form $D_j=\{d_j,\overline{d_j}\}$. The domain of $I_0$, say $D_0$, contains $(k+1)(n-1)+1$ alternatives. In particular, it contains an alternative $b_p$ that corresponds to the designated vertex $p \in V$, and for every vertex $v\in V\setminus \{p\}$, there are $k+1$ alternatives $b_v^\ell$, for $\ell\in [k+1]$. Essentially, these are identical $k+1$ 'copies' encoding the selection of $v$ in $I_0$, and play a significant role in the reverse direction of the reduction. As for the voters, there are two types of voters, {\it edge voters} and {\it vertex voters}. There is one edge voter for every edge $(i,j) \in E$, with a dependency graph having one edge from $I_j$ to $I_0$, and voting as follows: 
\begin{itemize}
    \item For the issue $I_0$, she votes conditioned on $I_j$ for $\{d_j:b_i\}$ if $i=p$ or otherwise for $\{d_j:b_i^\ell\}, \forall \ell \in[k+1]$. 
    \item For all other issues she is satisfied with any alternative. 
\end{itemize} 
For every vertex other than $p$, we also have a block of $L$ identical voters, where it suffices to take $L=m+1$. Each voter in the $j$-th block, with $j\in V\setminus\{p\}$ has a dependency graph with 1 edge, from $I_0$ to $I_j$ and votes as follows:
\begin{itemize}
    \item For the issue $I_j$, she is satisfied with the combinations $\{b_j^\ell:d_j\}$ for any $\ell$. Also, if the value of $I_0$ differs from $b_j^\ell$, for any $\ell$, she is satisfied with any value on $I_j$. Hence, the only restriction is that when the value of $I_0$ comes from an alternative corresponding to vertex $j$, the voter can be satisfied with respect to $I_j$ only by $d_j$.
    \item For all other issues, she is satisfied with any alternative. 
\end{itemize}

In total, we have $m+(n-1)L$ voters. We also use $k$ as the quota parameter, and we suppose the controller wants to enforce the alternative $b_p$ at issue $I_0$. Clearly, for every voter $i$, $\Delta_i\leq 1$ in her dependency graph, and the global dependency graph is a star centered on $I_0$. The maximum domain cardinality is $\bigoh{(kn)} = \bigoh{(n^2)}$.

For a better view of the construction we comment on Figure \ref{proofimage}, which illustrates the reduction to \cdaone. In particular, it illustrates only a part of the construction that pertains to the vertices of a subgraph $G'$ of the initial graph $G$ given in the instance of {\textsc{mod}} (as shown in the upper-left part of the figure). The figure also depicts the voters' ballots (rightmost part of the figure) and the global dependency graph which emerges (lower-left part of the figure). To be more precise, the lower-left part shows the union of the dependency graphs of all voters, where both orientations are present for the edges shown. Hence, the global dependency graph is simply a star centered on $I_0$. The connections in the rightmost part of the figure represent acceptable pairs of alternatives by voters. 
More precisely, a dotted connection between the alternatives $d_j$ and $b_j^\ell$ for some $j$ and $\ell$, represents the conditional approval ballot $\{b_j^\ell:d_j\}$ of the block of the $L$ identical vertex voters that correspond to $v_j$ of $G'$. A solid connection between the alternatives $d_j$ and $b_i^\ell$ (resp. between $d_i$ and $b_p$) represents the conditional approval ballot $\{d_j:b_i^\ell\}$ (resp. $\{d_j:b_p\}$) of an edge voter corresponding to edge $(v_i,v_j)$ (resp. $(p,v_j)$) of $G'$.

\begin{figure*}[t]

\caption{Illustrative example of the reduction in the proof of Theorem \ref{thm:cdaone-largedomain} \label{proofimage}}

\hspace{-0.25cm}
\scalebox{0.89}{
\tikzset{every picture/.style={line width=0.75pt}} 
\begin{tikzpicture}[x=0.75pt,y=0.75pt,yscale=-1,xscale=1]

\draw  [dash pattern={on 0.84pt off 2.51pt}]  (313.37,120.91) -- (278.37,184.53) ;
\draw  [dash pattern={on 0.84pt off 2.51pt}]  (313.37,120.91) .. controls (264.37,172.53) and (237.37,204.53) .. (264.37,217.53) ;
\draw  [dash pattern={on 0.84pt off 2.51pt}]  (313.37,120.91) .. controls (277.25,168) and (196.25,166) .. (268.37,298.56) ;
\draw  [dash pattern={on 0.84pt off 2.51pt}]  (355.37,123.39) -- (352.37,185.53) ;
\draw  [dash pattern={on 0.84pt off 2.51pt}]  (355.37,123.39) .. controls (360.25,168) and (373.25,154) .. (356.37,212.53) ;
\draw  [dash pattern={on 0.84pt off 2.51pt}]  (355.37,123.39) .. controls (378.37,184.53) and (373.37,257.09) .. (360.37,283.53) ;
\draw  [dash pattern={on 0.84pt off 2.51pt}]  (452.37,125.87) -- (500.37,190.53) ;
\draw  [dash pattern={on 0.84pt off 2.51pt}]  (452.37,125.87) .. controls (498.25,159) and (537.25,178) .. (512.25,214) ;
\draw  [dash pattern={on 0.84pt off 2.51pt}]  (452.37,125.87) .. controls (502.25,153) and (568.25,160) .. (510.37,283.53) ;
\draw  [color={rgb, 255:red, 0; green, 0; blue, 0 }  ,draw opacity=0 ][fill={rgb, 255:red, 219; green, 219; blue, 219 }  ,fill opacity=0.32 ][dash pattern={on 0.84pt off 2.51pt}] (297.37,86.05) .. controls (297.37,83.4) and (299.52,81.25) .. (302.17,81.25) -- (316.57,81.25) .. controls (319.22,81.25) and (321.37,83.4) .. (321.37,86.05) -- (321.37,130.16) .. controls (321.37,132.81) and (319.22,134.96) .. (316.57,134.96) -- (302.17,134.96) .. controls (299.52,134.96) and (297.37,132.81) .. (297.37,130.16) -- cycle ;
\draw  [color={rgb, 255:red, 0; green, 0; blue, 0 }  ,draw opacity=0 ][fill={rgb, 255:red, 219; green, 219; blue, 219 }  ,fill opacity=0.32 ][dash pattern={on 0.84pt off 2.51pt}] (346.37,85.22) .. controls (346.37,82.57) and (348.52,80.42) .. (351.17,80.42) -- (365.57,80.42) .. controls (368.22,80.42) and (370.37,82.57) .. (370.37,85.22) -- (370.37,129.33) .. controls (370.37,131.98) and (368.22,134.13) .. (365.57,134.13) -- (351.17,134.13) .. controls (348.52,134.13) and (346.37,131.98) .. (346.37,129.33) -- cycle ;
\draw  [color={rgb, 255:red, 0; green, 0; blue, 0 }  ,draw opacity=0 ][fill={rgb, 255:red, 219; green, 219; blue, 219 }  ,fill opacity=0.32 ][dash pattern={on 0.84pt off 2.51pt}] (390.37,85.22) .. controls (390.37,82.57) and (392.52,80.42) .. (395.17,80.42) -- (409.57,80.42) .. controls (412.22,80.42) and (414.37,82.57) .. (414.37,85.22) -- (414.37,129.33) .. controls (414.37,131.98) and (412.22,134.13) .. (409.57,134.13) -- (395.17,134.13) .. controls (392.52,134.13) and (390.37,131.98) .. (390.37,129.33) -- cycle ;
\draw  [color={rgb, 255:red, 0; green, 0; blue, 0 }  ,draw opacity=0 ][fill={rgb, 255:red, 219; green, 219; blue, 219 }  ,fill opacity=0.32 ][dash pattern={on 0.84pt off 2.51pt}] (439.37,84.4) .. controls (439.37,81.74) and (441.52,79.6) .. (444.17,79.6) -- (458.57,79.6) .. controls (461.22,79.6) and (463.37,81.74) .. (463.37,84.4) -- (463.37,128.5) .. controls (463.37,131.15) and (461.22,133.3) .. (458.57,133.3) -- (444.17,133.3) .. controls (441.52,133.3) and (439.37,131.15) .. (439.37,128.5) -- cycle ;
\draw    (339.37,198.53) .. controls (307.37,197.71) and (312.37,169.66) .. (311.37,125.87) ;
\draw    (337.37,220.53) .. controls (305.37,219.71) and (312.37,169.66) .. (311.37,125.87) ;
\draw    (342.37,301.04) .. controls (310.37,300.21) and (288.37,225.02) .. (311.37,125.87) ;
\draw  [dash pattern={on 0.84pt off 2.51pt}]  (407.37,129.17) -- (425.37,185.53) ;
\draw  [dash pattern={on 0.84pt off 2.51pt}]  (407.37,129.17) .. controls (426.37,172.14) and (482.91,161.21) .. (433.37,221.53) ;
\draw  [dash pattern={on 0.84pt off 2.51pt}]  (407.37,129.17) .. controls (484.37,181.23) and (477.37,187.03) .. (424.37,284.53) ;
\draw    (55.37,104.17) -- (75.51,112.22) ;
\draw [shift={(77.37,112.96)}, rotate = 201.79] [color={rgb, 255:red, 0; green, 0; blue, 0 }  ][line width=0.75]    (10.93,-3.29) .. controls (6.95,-1.4) and (3.31,-0.3) .. (0,0) .. controls (3.31,0.3) and (6.95,1.4) .. (10.93,3.29)   ;
\draw    (133.37,121.53) -- (102.97,98.77) ;
\draw [shift={(101.37,97.57)}, rotate = 396.82] [color={rgb, 255:red, 0; green, 0; blue, 0 }  ][line width=0.75]    (10.93,-3.29) .. controls (6.95,-1.4) and (3.31,-0.3) .. (0,0) .. controls (3.31,0.3) and (6.95,1.4) .. (10.93,3.29)   ;
\draw    (133.37,121.53) -- (104.1,138.29) ;
\draw [shift={(102.37,139.29)}, rotate = 330.2] [color={rgb, 255:red, 0; green, 0; blue, 0 }  ][line width=0.75]    (10.93,-3.29) .. controls (6.95,-1.4) and (3.31,-0.3) .. (0,0) .. controls (3.31,0.3) and (6.95,1.4) .. (10.93,3.29)   ;
\draw    (362.37,200.53) .. controls (385.37,193.1) and (377.37,164.7) .. (401.37,128.35) ;
\draw    (361.37,227.53) .. controls (406.37,233.32) and (373.37,157.27) .. (401.37,128.35) ;
\draw    (368.37,301.87) .. controls (413.37,307.65) and (403.37,168.83) .. (401.37,128.35) ;
\draw    (55.37,104.17) -- (77.42,98.78) ;
\draw [shift={(79.37,98.31)}, rotate = 526.27] [color={rgb, 255:red, 0; green, 0; blue, 0 }  ][line width=0.75]    (10.93,-3.29) .. controls (6.95,-1.4) and (3.31,-0.3) .. (0,0) .. controls (3.31,0.3) and (6.95,1.4) .. (10.93,3.29)   ;
\draw    (563.37,250.53) .. controls (514.25,485) and (13.25,302) .. (297.37,130.16) ;
\draw    (573.25,223) .. controls (592.25,152) and (619.25,74) .. (457.37,121.57) ;
\draw  [color={rgb, 255:red, 0; green, 0; blue, 0 }  ,draw opacity=0.24 ] (23.25,79) -- (150.37,79) -- (150.37,181) -- (23.25,181) -- cycle ;
\draw  [color={rgb, 255:red, 0; green, 0; blue, 0 }  ,draw opacity=0 ][fill={rgb, 255:red, 0; green, 0; blue, 0 }  ,fill opacity=0.17 ] (152.07,125.42) -- (172.51,125.26) -- (172.47,120.78) -- (186.17,129.63) -- (172.62,138.69) -- (172.58,134.21) -- (152.14,134.38) -- cycle ;
\draw  [color={rgb, 255:red, 0; green, 0; blue, 0 }  ,draw opacity=0.24 ] (183.37,55) -- (603.25,55) -- (603.25,379.53) -- (183.37,379.53) -- cycle ;
\draw    (33.91,242.5) -- (73.59,302.5) ;
\draw [shift={(75.25,305)}, rotate = 236.51] [fill={rgb, 255:red, 0; green, 0; blue, 0 }  ][line width=0.08]  [draw opacity=0] (10.72,-5.15) -- (0,0) -- (10.72,5.15) -- (7.12,0) -- cycle    ;
\draw [shift={(32.25,240)}, rotate = 56.51] [fill={rgb, 255:red, 0; green, 0; blue, 0 }  ][line width=0.08]  [draw opacity=0] (10.72,-5.15) -- (0,0) -- (10.72,5.15) -- (7.12,0) -- cycle    ;
\draw    (69.88,241.93) -- (81.62,297.07) ;
\draw [shift={(82.25,300)}, rotate = 257.97] [fill={rgb, 255:red, 0; green, 0; blue, 0 }  ][line width=0.08]  [draw opacity=0] (10.72,-5.15) -- (0,0) -- (10.72,5.15) -- (7.12,0) -- cycle    ;
\draw [shift={(69.25,239)}, rotate = 77.97] [fill={rgb, 255:red, 0; green, 0; blue, 0 }  ][line width=0.08]  [draw opacity=0] (10.72,-5.15) -- (0,0) -- (10.72,5.15) -- (7.12,0) -- cycle    ;
\draw    (100.63,241.94) -- (88.87,298.06) ;
\draw [shift={(88.25,301)}, rotate = 281.84000000000003] [fill={rgb, 255:red, 0; green, 0; blue, 0 }  ][line width=0.08]  [draw opacity=0] (10.72,-5.15) -- (0,0) -- (10.72,5.15) -- (7.12,0) -- cycle    ;
\draw [shift={(101.25,239)}, rotate = 101.84] [fill={rgb, 255:red, 0; green, 0; blue, 0 }  ][line width=0.08]  [draw opacity=0] (10.72,-5.15) -- (0,0) -- (10.72,5.15) -- (7.12,0) -- cycle    ;
\draw    (131.7,245.57) -- (96.8,303.43) ;
\draw [shift={(95.25,306)}, rotate = 301.1] [fill={rgb, 255:red, 0; green, 0; blue, 0 }  ][line width=0.08]  [draw opacity=0] (10.72,-5.15) -- (0,0) -- (10.72,5.15) -- (7.12,0) -- cycle    ;
\draw [shift={(133.25,243)}, rotate = 121.1] [fill={rgb, 255:red, 0; green, 0; blue, 0 }  ][line width=0.08]  [draw opacity=0] (10.72,-5.15) -- (0,0) -- (10.72,5.15) -- (7.12,0) -- cycle    ;
\draw  [color={rgb, 255:red, 0; green, 0; blue, 0 }  ,draw opacity=0.24 ] (21.37,212.53) -- (149.25,212.53) -- (149.25,356) -- (21.37,356) -- cycle ;
\draw  [color={rgb, 255:red, 0; green, 0; blue, 0 }  ,draw opacity=0 ][fill={rgb, 255:red, 0; green, 0; blue, 0 }  ,fill opacity=0.17 ] (182.26,281.86) -- (161.83,282.32) -- (161.93,286.79) -- (148.11,278.14) -- (161.53,268.88) -- (161.63,273.36) -- (182.07,272.91) -- cycle ;
\draw  [color={rgb, 255:red, 0; green, 0; blue, 0 }  ,draw opacity=0.51 ][fill={rgb, 255:red, 0; green, 0; blue, 0 }  ,fill opacity=1 ][line width=0.75]  (276.25,242) .. controls (276.25,241.45) and (275.8,241) .. (275.25,241) .. controls (274.7,241) and (274.25,241.45) .. (274.25,242) .. controls (274.25,242.55) and (274.7,243) .. (275.25,243) .. controls (275.8,243) and (276.25,242.55) .. (276.25,242) -- cycle ;
\draw  [color={rgb, 255:red, 0; green, 0; blue, 0 }  ,draw opacity=0.51 ][fill={rgb, 255:red, 0; green, 0; blue, 0 }  ,fill opacity=1 ][line width=0.75]  (276.25,271) .. controls (276.25,270.45) and (275.8,270) .. (275.25,270) .. controls (274.7,270) and (274.25,270.45) .. (274.25,271) .. controls (274.25,271.55) and (274.7,272) .. (275.25,272) .. controls (275.8,272) and (276.25,271.55) .. (276.25,271) -- cycle ;
\draw  [color={rgb, 255:red, 0; green, 0; blue, 0 }  ,draw opacity=0.51 ][fill={rgb, 255:red, 0; green, 0; blue, 0 }  ,fill opacity=1 ][line width=0.75]  (276.25,256) .. controls (276.25,255.45) and (275.8,255) .. (275.25,255) .. controls (274.7,255) and (274.25,255.45) .. (274.25,256) .. controls (274.25,256.55) and (274.7,257) .. (275.25,257) .. controls (275.8,257) and (276.25,256.55) .. (276.25,256) -- cycle ;
\draw  [color={rgb, 255:red, 0; green, 0; blue, 0 }  ,draw opacity=0 ][fill={rgb, 255:red, 219; green, 219; blue, 219 }  ,fill opacity=0.32 ][dash pattern={on 4.5pt off 4.5pt}] (187.37,204.73) .. controls (187.37,188.16) and (200.8,174.73) .. (217.37,174.73) -- (564.37,174.73) .. controls (580.94,174.73) and (594.37,188.16) .. (594.37,204.73) -- (594.37,294.73) .. controls (594.37,311.3) and (580.94,324.73) .. (564.37,324.73) -- (217.37,324.73) .. controls (200.8,324.73) and (187.37,311.3) .. (187.37,294.73) -- cycle ;
\draw  [color={rgb, 255:red, 0; green, 0; blue, 0 }  ,draw opacity=0.51 ][fill={rgb, 255:red, 0; green, 0; blue, 0 }  ,fill opacity=1 ][line width=0.75]  (353.25,243) .. controls (353.25,242.45) and (352.8,242) .. (352.25,242) .. controls (351.7,242) and (351.25,242.45) .. (351.25,243) .. controls (351.25,243.55) and (351.7,244) .. (352.25,244) .. controls (352.8,244) and (353.25,243.55) .. (353.25,243) -- cycle ;
\draw  [color={rgb, 255:red, 0; green, 0; blue, 0 }  ,draw opacity=0.51 ][fill={rgb, 255:red, 0; green, 0; blue, 0 }  ,fill opacity=1 ][line width=0.75]  (353.25,272) .. controls (353.25,271.45) and (352.8,271) .. (352.25,271) .. controls (351.7,271) and (351.25,271.45) .. (351.25,272) .. controls (351.25,272.55) and (351.7,273) .. (352.25,273) .. controls (352.8,273) and (353.25,272.55) .. (353.25,272) -- cycle ;
\draw  [color={rgb, 255:red, 0; green, 0; blue, 0 }  ,draw opacity=0.51 ][fill={rgb, 255:red, 0; green, 0; blue, 0 }  ,fill opacity=1 ][line width=0.75]  (353.25,257) .. controls (353.25,256.45) and (352.8,256) .. (352.25,256) .. controls (351.7,256) and (351.25,256.45) .. (351.25,257) .. controls (351.25,257.55) and (351.7,258) .. (352.25,258) .. controls (352.8,258) and (353.25,257.55) .. (353.25,257) -- cycle ;
\draw  [color={rgb, 255:red, 0; green, 0; blue, 0 }  ,draw opacity=0.51 ][fill={rgb, 255:red, 0; green, 0; blue, 0 }  ,fill opacity=1 ][line width=0.75]  (424.25,242) .. controls (424.25,241.45) and (423.8,241) .. (423.25,241) .. controls (422.7,241) and (422.25,241.45) .. (422.25,242) .. controls (422.25,242.55) and (422.7,243) .. (423.25,243) .. controls (423.8,243) and (424.25,242.55) .. (424.25,242) -- cycle ;
\draw  [color={rgb, 255:red, 0; green, 0; blue, 0 }  ,draw opacity=0.51 ][fill={rgb, 255:red, 0; green, 0; blue, 0 }  ,fill opacity=1 ][line width=0.75]  (424.25,271) .. controls (424.25,270.45) and (423.8,270) .. (423.25,270) .. controls (422.7,270) and (422.25,270.45) .. (422.25,271) .. controls (422.25,271.55) and (422.7,272) .. (423.25,272) .. controls (423.8,272) and (424.25,271.55) .. (424.25,271) -- cycle ;
\draw  [color={rgb, 255:red, 0; green, 0; blue, 0 }  ,draw opacity=0.51 ][fill={rgb, 255:red, 0; green, 0; blue, 0 }  ,fill opacity=1 ][line width=0.75]  (424.25,256) .. controls (424.25,255.45) and (423.8,255) .. (423.25,255) .. controls (422.7,255) and (422.25,255.45) .. (422.25,256) .. controls (422.25,256.55) and (422.7,257) .. (423.25,257) .. controls (423.8,257) and (424.25,256.55) .. (424.25,256) -- cycle ;
\draw  [color={rgb, 255:red, 0; green, 0; blue, 0 }  ,draw opacity=0.51 ][fill={rgb, 255:red, 0; green, 0; blue, 0 }  ,fill opacity=1 ][line width=0.75]  (503.25,243) .. controls (503.25,242.45) and (502.8,242) .. (502.25,242) .. controls (501.7,242) and (501.25,242.45) .. (501.25,243) .. controls (501.25,243.55) and (501.7,244) .. (502.25,244) .. controls (502.8,244) and (503.25,243.55) .. (503.25,243) -- cycle ;
\draw  [color={rgb, 255:red, 0; green, 0; blue, 0 }  ,draw opacity=0.51 ][fill={rgb, 255:red, 0; green, 0; blue, 0 }  ,fill opacity=1 ][line width=0.75]  (503.25,272) .. controls (503.25,271.45) and (502.8,271) .. (502.25,271) .. controls (501.7,271) and (501.25,271.45) .. (501.25,272) .. controls (501.25,272.55) and (501.7,273) .. (502.25,273) .. controls (502.8,273) and (503.25,272.55) .. (503.25,272) -- cycle ;
\draw  [color={rgb, 255:red, 0; green, 0; blue, 0 }  ,draw opacity=0.51 ][fill={rgb, 255:red, 0; green, 0; blue, 0 }  ,fill opacity=1 ][line width=0.75]  (503.25,257) .. controls (503.25,256.45) and (502.8,256) .. (502.25,256) .. controls (501.7,256) and (501.25,256.45) .. (501.25,257) .. controls (501.25,257.55) and (501.7,258) .. (502.25,258) .. controls (502.8,258) and (503.25,257.55) .. (503.25,257) -- cycle ;
\draw  [fill={rgb, 255:red, 0; green, 0; blue, 0 }  ,fill opacity=0.29 ] (27.25,97.4) .. controls (27.25,89.45) and (33.7,83) .. (41.65,83) -- (148.25,83) .. controls (148.25,83) and (148.25,83) .. (148.25,83) -- (148.25,140.6) .. controls (148.25,148.55) and (141.8,155) .. (133.85,155) -- (27.25,155) .. controls (27.25,155) and (27.25,155) .. (27.25,155) -- cycle ;

\draw (266,185.43) node [anchor=north west][inner sep=0.75pt]    {$b^{1}_{1}$};
\draw (266,212.43) node [anchor=north west][inner sep=0.75pt]    {$b^{2}_{1}$};
\draw (271,285.43) node [anchor=north west][inner sep=0.75pt]    {$b^{k+1}_{1}$};
\draw (342,187.4) node [anchor=north west][inner sep=0.75pt]    {$b^{1}_{2}$};
\draw (342,214.4) node [anchor=north west][inner sep=0.75pt]    {$b^{2}_{2}$};
\draw (347,287.4) node [anchor=north west][inner sep=0.75pt]    {$b^{k+1}_{2}$};
\draw (494,187.4) node [anchor=north west][inner sep=0.75pt]    {$b^{1}_{4}$};
\draw (495,214.4) node [anchor=north west][inner sep=0.75pt]    {$b^{2}_{4}$};
\draw (494,289.4) node [anchor=north west][inner sep=0.75pt]    {$b^{k+1}_{4}$};
\draw (560,227.4) node [anchor=north west][inner sep=0.75pt]    {$b_{p}$};
\draw (304,58.02) node [anchor=north west][inner sep=0.75pt]    {$I_{1}$};
\draw (348,58.85) node [anchor=north west][inner sep=0.75pt]    {$I_{2}$};
\draw (397,59.68) node [anchor=north west][inner sep=0.75pt]    {$I_{3}$};
\draw (444,58.85) node [anchor=north west][inner sep=0.75pt]    {$I_{4}$};
\draw (391,328.43) node [anchor=north west][inner sep=0.75pt]    {$I_{0}$};
\draw (419,185.67) node [anchor=north west][inner sep=0.75pt]    {$b^{1}_{3}$};
\draw (417,212.67) node [anchor=north west][inner sep=0.75pt]    {$b^{2}_{3}$};
\draw (412,286.67) node [anchor=north west][inner sep=0.75pt]    {$b^{k+1}_{3}$};
\draw (392.37,81.62) node [anchor=north west][inner sep=0.75pt]    {$\bar{ \begin{array}{l}
d_{3}\\
d_{3}\\
\end{array}}$};
\draw (441.37,80.79) node [anchor=north west][inner sep=0.75pt]    {$\bar{ \begin{array}{l}
d_{4}\\
d_{4}\\
\end{array}}$};
\draw (349.17,81.79) node [anchor=north west][inner sep=0.75pt]    {$\bar{ \begin{array}{l}
d_{2}\\
d_{2}\\
\end{array}}$};
\draw (300.17,81.79) node [anchor=north west][inner sep=0.75pt]    {$\bar{ \begin{array}{l}
d_{1}\\
d_{1}\\
\end{array}}$};
\draw (82,86.99) node [anchor=north west][inner sep=0.75pt]    {$v_{1}$};
\draw (36,92.32) node [anchor=north west][inner sep=0.75pt]    {$v_{2}$};
\draw (81,106.78) node [anchor=north west][inner sep=0.75pt]    {$v_{3}$};
\draw (135,110.01) node [anchor=north west][inner sep=0.75pt]    {$p$};
\draw (82,129.37) node [anchor=north west][inner sep=0.75pt]    {$v_{4}$};
\draw (187,362.23) node [anchor=north west][inner sep=0.75pt]    {voters' preferences};
\draw (79,307.43) node [anchor=north west][inner sep=0.75pt]    {$I_{0}$};
\draw (27,215.02) node [anchor=north west][inner sep=0.75pt]    {$I_{1}$};
\draw (60,215.85) node [anchor=north west][inner sep=0.75pt]    {$I_{2}$};
\draw (94,216.68) node [anchor=north west][inner sep=0.75pt]    {$I_{3}$};
\draw (129,216.85) node [anchor=north west][inner sep=0.75pt]    {$I_{4}$};
\draw (26,338) node [anchor=north west][inner sep=0.75pt]   [align=left] {\footnotesize{union of dep. graphs}};
\draw (29,161.4) node [anchor=north west][inner sep=0.75pt]    {$G'$ subgraph of $G$};

\end{tikzpicture}}

\vspace{-3cm}
\end{figure*}

Suppose there exists a set $S$ of vertices in $G$ of size at most $k$, say without loss of generality that $S = \{1,\dots,k\} \subseteq V$, whose deletion leaves $p$ as the only vertex of maximum out-degree. 
We now choose to delete the corresponding alternatives $\{d_1,\dots,d_k\}$ from the issues $\{I_1,\dots,I_k\}$. If we select $b_p$ for the issue $I_0$, then the total dissatisfaction score can be brought down to $m-deg_{V\setminus S}(p)$ by choosing $d_j$ for every issue $I_j$ where $d_j$ has not been deleted. To see this, the only edge voters that are satisfied with respect to $I_0$ are voters corresponding to edges that are outgoing from $p$ and whose other endpoint belongs to $V\setminus S$. Hence all remaining $m- deg_{V\setminus S}(p)$ voters will be dissatisfied with respect to $I_0$. Regarding the vertex voters, they will all be satisfied on all issues.

On the other hand, if we select for $I_0$ some $b_j^\ell$ for any $\ell\in [k+1]$, we need to consider two cases, depending on $j$. If $j\in V\setminus S$, then by the same reasoning as before, the best we could achieve is to 
have a dissatisfaction score equal to $m- deg_{V\setminus S}(j)$. But since $p$ has the maximum out-degree, this would yield a worse solution. Now suppose $j\in S$. Then we know that $d_j$ has been deleted from $I_j$. Hence, the $j$-th block of vertex voters will be dissatisfied with respect to $I_j$, and since $L>m$, this cannot yield an optimal solution.  
To conclude, after the deletion of the selected alternatives, $b_p$ has to be selected for $I_0$ in any optimal solution. 

For the reverse direction, suppose that there is a set $D''$ of at most $k$ alternatives, the deletion of which, forces $b_p$ to be selected for $I_0$ in every optimal solution. It is without loss of generality to assume that $D''$ does not contain anything from $D_0$. To elaborate on this claim, since there are $k+1$ copies of alternatives for every $i\in V\setminus \{p\}$ that have an identical role, there is no change in the optimal outcome by deleting up to $k$ alternatives from $I_0$ (some alternative will survive for every $i$). 
Moreover, we can assume that none of the deleted alternatives equals $\overline{d_j}$ for some issue $I_j\neq I_0$ since if it were, we can swap it with $d_j$ without harming the cost of the optimal solution (one cannot strengthen the support of $b_p$ in $I_0$ by deleting $\overline{d_j}$ for some $j$). Also, bear in mind that we are not allowed to delete both $d_j$ and $\overline{d_j}$ from an issue $I_j, j\in [n]$, as there are no other choices left for $I_j$. 

To summarize, the deleted alternatives must come from distinct issues among $I_1,\dots, I_n$ and they all correspond to some $d_j$ for $j\in [n]$. It is now easy to observe that deleting from $V$ the set $S$ formed by the vertices corresponding to these alternatives in $D''$, makes $p$ the unique vertex of maximum out-degree in the induced subgraph of $G$. If not, there is a vertex, say $v\in V\setminus S$, with greater or equal out-degree. In that case, if we select $b_v^{\ell'}$ for $I_0$ for some arbitrary $\ell'$, and $d_j$ for all issues $I_j$, for which $d_j$ has not been deleted, we will obtain a solution with at most the same dissatisfaction score as the optimal solution that used $b_p$. Indeed, we will have fewer or equal dissatisfactions from the edge voters with respect to $I_0$, and also all the blocks of the vertex voters will be satisfied (the block of voters who care about $I_v$ is satisfied because $d_v$ has not been deleted, since $v\in V\setminus S$). This contradicts the fact that $b_p$ was elected for $I_0$ in every optimal solution.

 For the \np-hardness of \caaone, the proof is based on a similar reasoning as in the proof of \cdaone, but with appropriate adjustments. First, it is more convenient to perform a reduction from a slightly different problem, which is the {\textsc{max-outdegree addition (moa)}} problem defined and proven to be \np-hard by \citeauthor{BU09} \citeyear{BU09}.

\vspace{0.25cm}

\noindent\fbox{%
    \parbox{0.974\columnwidth}{
\textbf{Instance:} A directed graph $G=(V_1\cup V_2,E)$, where $V_1$ denotes the set of registered vertices, and $V_2$ is the set of unregistered vertices, a distinguished vertex $p\in V_1$ and an integer $k\geq 1$.\\
\textbf{Output:} Does there exist a set $V'\subseteq V_2$ with $|V'|\leq k$ such that $p$ is the only vertex that has maximum outdegree in $G[V_1 \cup V']$?
}}

\vspace{0.25cm}

Starting from an instance of {\textsc{moa}}, where $n = |V_1|+|V_2|$, let $I= \{I_0, I_1,I_2,\dots,I_n\}$. For $j\in V_1$, we have two qualified alternatives, $D_j=\{d_j,\overline{d_j}\}$ and no spoiler ones. For $j\in V_2$, we have one qualified alternative\footnote{If one wishes to avoid issues with unary starting domain, we can also add one dummy qualified alternative, so that no issue is trivialized before the addition of any spoiler alternatives.}, $D_j=\{\overline{d_j}\}$, and we will have $d_j$ as a spoiler alternative, $D_j' = \{d_j\}$. The domain of $I_0$ corresponds to all the vertices and equals $D_0 = \{b_1,\dots, b_n\}$. In contrast to \cdvone, we do not need to have $k+1$ ``copies" for each $b_i$, since the spoiler alternatives that will be added are not going to be from $D_0$. 
As for the voters, there is one edge voter for every edge of the graph, regardless of whether its endpoints belong to $V_1$ or $V_2$ and one vertex voter for every vertex of the graph. All voters have similar preferences as in the \cdaone\ reduction, from which their ballots for each issue $I_j$ with $j \in \{0,1,\dots,n\}$ can be immediately obtained by replacing, $\{b_j^\ell\}_{\forall \ell \in [k+1]}$ with $b_j$. For example, an edge voter arising from an edge $(i, j)$ will vote for the combination $\{d_j:b_i\}$ regarding $I_0$. Using similar arguments as in the proof for \cdaone, we conclude that there is a way to add up to $k$ vertices and make $p$ the unique vertex with maximum out-degree if and only if there is a set of at most $k$ alternatives to add in the \caaone\ instance to fulfill controller's will.
\end{proof}

Notably, moving to a constant domain size, the considered problems, \cdaone\ and \caaone, seem to behave differently, as the following result indicates.

\begin{proposition}
\label{prop:cdaone}
\cdaone\ can be solved in polynomial time, when $\Delta\leq 1$, the treewidth of the global dependency graph is constant and the domain size is also constant for every issue.
\end{proposition}
\begin{proof}
Let $q$ be the quota parameter and let $I_j$ be the issue where the controller wants to enforce a specific alternative. If $q\geq |D_j|-1$, then we can simply delete precisely all other $|D_j|-1$ alternatives of $I_j$ so that the controller's will is the only choice left. If $q< |D_j|-1$, this implies that $q = \bigoh{(1)}$. Then we can check all possible ways of picking up to $q$ items from the available set of all alternatives of all issues (a polynomial in $m$). For every such combination, and since the conditions of Theorem \ref{thm:tw} hold, we can solve the remaining \cms\ instance and check if we can have the controller's choice in every optimal solution, by solving \cms\ with and without the designated alternative. 
\end{proof}

Hence, a constant domain size makes a difference for \cdaone\ when we stick to the assumptions from Section \ref{sec:opt} on each $\Delta_i$ and on the treewidth. For \caaone, we are not yet aware if the same result holds (the proof arguments certainly do not go through), and we leave this as an interesting open problem. However, we have established intractability, as soon as we move to slightly richer instances with $\Delta_i\leq 2$. 

\begin{theorem}
\caaone\ is \np-hard, when $\Delta\leq 2$, even when the treewidth of the global dependency graph is at most one and even for a binary domain size in every issue. 
\end{theorem}
\begin{proof}
Consider an instance $P$ of \textsc{vertex cover}, asking if there is a cover of size at most $k$ in a graph $G = (V, E)$, with $|V|=n$, $|E|=m$. We create an instance $P'$ of \caaone\ with $n+1$ issues $I = \{I_0, I_1,\dots, I_n\}$. The issue $I_0$ has two qualified alternatives, $D_0 = \{d_0, \overline{d_0}\}$. Each issue $I_j$ for $j\in [n]$ corresponds to a vertex of $G$, and has one qualified alternative, denoted by $\overline{d_j}$, and one unqualified one denoted by $d_j$. 
Formally, $D_j = \{\overline{d_j}\}$ and $D'_j = \{d_j\}$, for $j\in [n]$. As for the voters, we have a total of $2m-1$ voters. The first $m$ voters correspond to the edges of $G$, and they are satisfied with all the alternatives in the issues $I_j$, $j\in [n]$. For issue $I_0$, each edge voter has a dependence on the two issues corresponding to its endpoints. In particular, for an edge $(j, \ell)$, the corresponding edge voter has a dependence of $I_0$ on both $I_j$ and $I_\ell$. He is satisfied with respect to $I_0$, only when either $d_j$ or $d_\ell$ is selected, and $d_0$ is selected as well. Thus he is satisfied with the combinations $\{(d_j,x):d_0\}$ for any $x\in \{d_\ell, \overline{d_\ell}\}$, and with $\{(x,d_\ell):d_0\}$ for any $x\in \{d_j, \overline{d_j}\}$. 
These together encode precisely the constraint $(d_j \vee d_\ell):d_0$. Any other combination of alternatives of $I_j, I_\ell$, and $I_0$ make this edge voter dissatisfied with respect to $I_0$. The remaining $m-1$ dummy voters are satisfied with all the alternatives of the first $n$ issues and are also satisfied only with $\overline{d_0}$ for issue $I_0$. To complete the construction, we use the parameter $k$ appearing in the instance $P$ as the quota of $P'$, and we assume that the controller wants to enforce $d_0$ on issue $I_0$. It is easy to check that the maximum in-degree for every voter is at most two, and that the global dependency graph is a star centered on $I_0$, and hence with treewidth equal to one.

Suppose that $P$ has a vertex cover $S$ of size at most $k$. We then add in $P'$ the unqualified alternatives for the issues that belong to the vertex cover of $G$. By selecting those alternatives, and with $d_0$ for $I_0$, and any alternative for the remaining issues, we claim that all the edge voters are satisfied with respect to $I_0$ (since for every edge, at least one of the added alternatives together with $d_0$ satisfy the corresponding constraint). Thus, there is only 1 unit of dissatisfaction from every dummy voter on $I_0$, with a total score of $m-1$. Any solution where $d_0$ is not the selected choice for $I_0$ would dissatisfy all the edge voters, and would have a score of at least $m$, hence cannot be optimal. Thus, we have ensured that in every optimal solution, $I_0$ is assigned the value of $d_0$.   

For the reverse direction, suppose that there is a set of at most $k$ unqualified alternatives that, when added, ensure that $d_0$ is selected in every optimal solution. We know that selecting $d_0$ causes the dummy voters to be dissatisfied, hence the optimal dissatisfaction score is at least $m-1$. If $\overline{d_0}$ was chosen for $I_0$, we know that the total dissatisfaction score is $m$ (due to the edge voters), and since this cannot be optimal, we have that the dissatisfaction score in an optimal solution is exactly $m-1$. But this means that all the remaining $m$ voters, or equivalently all edge voters, have to be satisfied with all issues in the optimal solution, i.e., satisfied with $I_0$ as well. Thus, the added alternatives need to satisfy every edge voter, which means that if a voter's dependence of $I_0$ is based on issues $I_j$ and $I_\ell$, then either $d_j$ or $d_\ell$ has been added (or both), and hence the set of added alternatives correspond to a vertex cover of size at most $k$. 
\end{proof}

\raggedbottom

Overall, we end this subsection by concluding that \cms\ is indeed sufficiently robust against malicious actions, in most of the variants of the control problem considered. The behavior regarding the addition or deletion of alternatives does exhibit differences with that of deleting or adding voters. In some cases, such as in adding alternatives, we get the stronger guarantee of immunity, compared to \np-hardness. On the other hand, when deleting alternatives, we have vulnerability in more settings than when deleting voters. Still, we do have computational hardness, when the domain size is large enough.


\section{Conclusions and Further Work}

Our work is centered around the \cms\ rule, a relatively new and highly natural voting rule for expressing preferential dependencies with approval-based conditional ballots in elections over multiple interdependent issues. We focused on computational aspects of \cms\ elections, from the perspective of the winner determination problem using exact (polynomial and parameterized) and approximate algorithms, as well as from the perspective of strategic attempts to control election outcomes. 
We conclude that \cms\ provides a satisfactory tradeoff between expressiveness and efficiency under certain assumptions and at the same time exhibits sufficient resilience against control actions in the considered settings, which makes it a strong candidate for real-life multi-issue elections.

There are still several interesting open problems and even broader directions for future research.
It is conceivable that approximation guarantees can be obtained for instances with higher expressiveness (i.e., higher in-degrees) than those considered in Section \ref{sec:approx}. Additionally, one can also consider other objective functions, such as the Conditional Minimax rule \cite{BL16}, for which, algorithmic results remain elusive. In principle, one can take any other voting rule defined for approval ballots and explore potential generalizations in the setting with conditional approval ballots, as done for instance by \citeauthor{brill2023proportionality} \citeyear{brill2023proportionality} with the Proportional Approval Voting rule (PAV) and the Method of Equal Shares (MES). 
Furthermore, from a strategic point of view, there is one case of our control questions that has been left open, namely, the complexity of \caaone, under a binary domain size and with $\Delta_i\leq 1$. But even more interestingly, one can study several other relevant questions on controlling elections. For instance, destructive versions of control, control of more than one (but not all of the) issues, control by deleting entire issues (rather than alternatives of issues) are naturally motivated. Moreover, we could incorporate the notion of approximation into control problems, so as to demand that the controller's will should be satisfied in every approximate solution of (hard) \cms\ instances. Yet another direction is to go beyond the worst-case complexity of control problems. Another open problem concerning strategic moves has to do with incentive compatibility. It has been proven that \cms\ is not strategyproof \cite{BL16}, but the complexity of manipulation has not been examined yet. Finally, as a highly critical area of future work, we emphasize the importance of obtaining real or synthetic data on elections over interdependent issues (currently nonexistent in public preference data libraries \cite{preflib}) and of generating experimental results that complement our theoretical findings.


\section*{Acknowledgements} We are grateful to the anonymous IJCAI '21 reviewer who pointed out to us the generalization of Theorem \ref{thm:tw} to Theorem \ref{thm:twgen}. We are also thankful to all the reviewers of IJCAI '20, IJCAI '21 and TCS for their constructive feedback. This research was supported by the Hellenic Foundation for Research and Innovation (HFRI), by the “1st Call for HFRI Research Projects to support faculty members
and researchers and the procurement of high-cost research equipment”
(Project Num. HFRI-FM17-3512).
The second author was also partially supported by the ``3rd Call for HFRI PhD Fellowships" (Fellowship Number: 5163).
He was also supported by the European Union
(ERC, PRO-DEMOCRATIC, 101076570). Views and opinions expressed are however those of the
author only and do not necessarily reflect those of the European Union or the European Research
Council. Neither the European Union nor the granting authority can be held responsible for them.

\begin{center}
  \includegraphics[width=4cm]{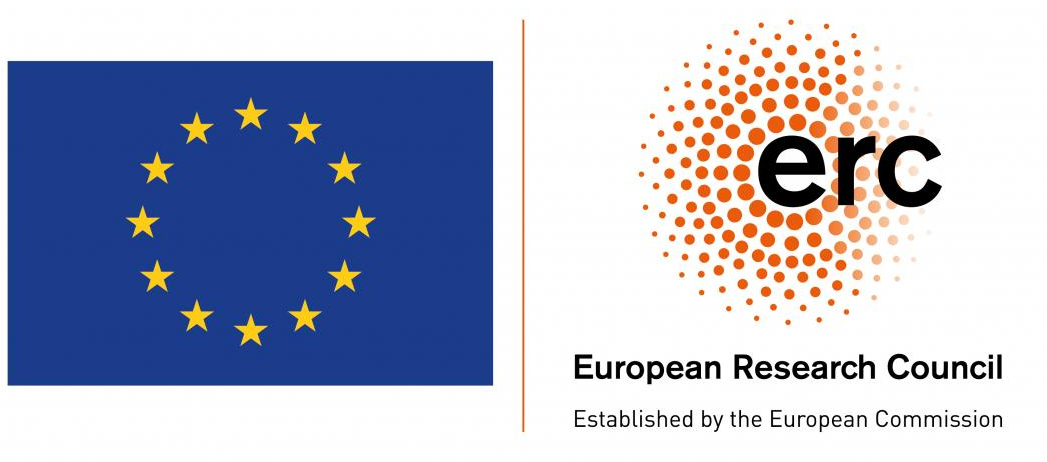}
\end{center}

\vskip 0.2in
\bibliography{references}
\bibliographystyle{theapa}

\end{document}